\documentclass[prl,twocolumn,superscriptaddress%,longbibliography
]{revtex4-1}

\usepackage{epsfig} % Allows the inclusion of eps files
\usepackage{epic} % Enhanced picture mode
\usepackage{eepic} % Extensions for epic
\usepackage{url} % URL handling
\usepackage{longtable} % Tables that continue onto multiple pages
\usepackage{mathrsfs} % Support for \mathscr script
\usepackage{multirow} % Span rows in tables
\usepackage{bigstrut} % Space struts in tables up and down
\usepackage{amssymb} % AMS math symbols and helpers
\usepackage{graphicx} % Enhanced graphics support
\usepackage{setspace} % Adjust spacing in captions, single by default
\usepackage{xspace} % Automatically adjusting space after macros
\usepackage{amsmath} % \text, and other math formatting options
\usepackage{siunitx} % \num{} formatting and SI unit formatting
\usepackage{booktabs} % Enhanced tables with \Trule, etc.
\usepackage[
    colorlinks=true,
    linkcolor=blue,
    citecolor=blue,
    urlcolor=blue,
    hypertexnames=false
]{hyperref}
\usepackage[noabbrev]{cleveref} % Automatically determine \cref type
\usepackage{bbold} % \mathbb{1}

\usepackage{subfigure}
\usepackage{braket}
\usepackage{extarrows} % arrows \xlongrigharrow{}
\usepackage{amsfonts}
\usepackage{indentfirst}
\usepackage{amsthm} % math
\usepackage{soul}
\usepackage{color}
\definecolor{darkblue}{rgb}{0,0.02,0.45}
\definecolor{darkred}{rgb}{0.45,0.02,0} 

\newcommand{\fsz}{\small}

\newcommand{\ud}{\mathrm{d}}

\newcommand{\lefta}{\left\langle}
\newcommand{\righta}{\right\rangle}
\newcommand{\leftr}{\left[}
\newcommand{\rightr}{\right]}
\newcommand{\nn}{\nonumber}
\newcommand{\one}{\mathbb{1}}
\newcommand{\T}{{\mkern-1.5mu\mathsf{T}}}

\usepackage{etoolbox}
 
\newtheorem{myTheo}{Theorem}
\AfterEndEnvironment{myTheo}{\noindent\ignorespaces}

\newcommand{\RNum}[1]{\uppercase\expandafter{\romannumeral #1\relax}}
\newcommand{\tx}[1]{\textmd{#1}}
\newcommand{\onefrac}[1]{\frac{1}{#1}}
\newcommand{\refeq}[1]{\textmd{Eq.\ }(\ref{#1})}
\newcommand{\reffg}[1]{\textmd{Fig.\ }\ref{#1}}

\newcommand{\vrr}{\mathbf{r}}

\newcommand{\mR}{\mathcal{R}}
\newcommand{\mI}{\mathcal{I}}
\newcommand{\mD}{\mathcal{D}}
\newcommand{\mP}{\mathcal{P}}
\newcommand{\Tr}{\operatorname{Tr}}
\newcommand{\im}{\operatorname{Im}}
\newcommand{\re}{\operatorname{Re}}

\allowdisplaybreaks
\def\em{\it}

\begin{document}
%\title{Raman scattering of phonon - Kitaev spin liquids coupled system at finite temperature and field with an application on $\alpha$-$\tx{RuCl}_3$}

\title {Footprints  of the Kitaev spin liquid in the Fano  lineshape  of  the  Raman  active  optical  phonons} 
\author{ Kexin Feng}
\affiliation{School of Physics and Astronomy, University of Minnesota, Minneapolis, MN 55455, USA}
\author{Swetlana Swarup}
\affiliation{School of Physics and Astronomy, University of Minnesota, Minneapolis, MN 55455, USA}
\author{Natalia B. Perkins}
\affiliation{School of Physics and Astronomy, University of Minnesota, Minneapolis, MN 55455, USA}

\date{\today}
\begin{abstract}

 We develop a  theoretical  description of the Raman spectroscopy in the spin-phonon coupled Kitaev system and  show that it can provide 
  observable signatures
  of  fractionalized excitations
  characteristic of the underlying spin liquid phase.
  In particular, we obtain the explicit form of the phonon modes and construct the coupling Hamiltonian based on the $D_{3d}$ symmetry. 
  We then systematically compute the Raman intensity and  show that the spin-phonon coupling renormalizes phonon propagators and generates the salient Fano linshape.
  We find that the temperature evolution of the Fano lineshape
  displays two crossovers, and the low temperature crossover shows pronounced magnetic field dependence. 
  We thus identify the observable effect of the Majorana fermions and the
 $Z_2$ gauge fluxes encoded in the Fano lineshape. Our results  are consistent with the phonon Raman scattering experiments in the candidate  material $\alpha\tx{-RuCl}_3$.
\end{abstract}
\maketitle
 
 {\em Introduction.--}  Raman spectroscopy  has proven to be a sensitive experimental probe to study the ground state properties and the dynamics of various strongly correlated systems \cite{Devereaux2007}. For magnetic insulators, Raman process couples to the dynamically induced electron-hole pair, that connects to the low-energy magnetic states. 
 In magnetically ordered states, the magnetic Raman response shows  polarization-dependent peak structure,
  arising  predominantly  from one- and two-magnon excitations \cite{Fleury1968,Shastry1990,Chubukov1995,Perkins2008,Perkins2013,Yang2021}. 
 In quantum spin liquid (QSL) phase,
%  where simple spin wave picture of the low energy excitations is not applicable,
 the Raman spectrum of such low-energy states reveals characteristic low-energy continua, which are fundamentally different from the 
  dispersive collective modes in 
  ordered states. These continua reflect 
%   the statistics of the excitations and unveil
  the fractionalization of spins, a hallmark of QSL \cite{Ko2010,Knolle2014,Perreault2015,Brent2016-short,Brent2016-long,Nasu2016,Rousochatzakis2019,Fu2017,Metavitsiadis2021}.
  
Recently, significant efforts have been made in the investigation of QSL state of matter. Mott insulators  with  strong  spin-orbit  coupling,  e.g $\alpha$-$\tx{RuCl}_3$ \cite{Plumb2014,Sears2015,Banerjee2016,Banerjee2017,banerjee2018excitations,Little2017,Sandilands2015,Li2019,Wulferding2020,Sahasrabudhe2020,Lin2020,Wang2020}, are promising to realize Kitaev QSL. This QSL is motivated by the famous Kitaev spin model with bond-dependent Ising interactions on a two-dimensional honeycomb lattice \cite{Kitaev2006}. It is exactly solvable with known gapless QSL ground state. In this model, the spins fractionalize into static $Z_2$ gauge fluxes and itinerant Majorana fermions amenable to experimental detection.

While various dynamical probes \cite{Knolle2014,Knolle2014a,Knolle2015,Gabor2016,Halasz2019,Rousochatzakis2019,Wan2019} have been exploited in several materials to look for signatures of spin fractionalization and their proximity to the Kitaev QSL, employing phonon dynamics and the spin-lattice coupling to detect Kitaev QSL is less investigated. It was recently suggested that sound attenuation from the phonon decaying into a pair of  Majorana  fermions \cite{Metavitsiadis2020,Ye2020,Feng2021}  and  the  Hall  viscosity induced by  time-reversal breaking spin Hamiltonian \cite{Ye2020,Feng2021}
may potentially serve as such probe.
The importance of the spin-phonon coupling in the Kitaev materials is also shown in the interpretation of the  thermal Hall transport measurements
~\cite{Kasahara2018,Ye2018,Vinkler2018}.
 
In this letter, we focus on the Raman spectroscopy of optical phonons, and particularly the salient Fano line shape, which arises when the phonon resonance peak couples to the magnetic continuum \cite{Fano1961}. This effect is attributed to {\it spin-dependent} electron polarizability \cite{Suzuki1973, Moriya1967}, which involves a microscopic description of both spin-photon coupling and spin-phonon couplings.
A recent work Ref.\ \cite{Metavitsiadis2021} shows that even the simplest form of the couplings can give rise to the Fano line shape. 
In the experimental studies of  the candidate material $\alpha$-$\tx{RuCl}_3$ \cite{Sandilands2015,Glamazda2017, Li2019,Wulferding2020,Sahasrabudhe2020,Lin2020}, the pronounced temperature  and field dependence of Fano lineshape indicate rich information about the underlying spin liquid phase that awaits exploration.
However, up to now a clear theoretical description of the Raman scattering in a Kitaev spin-phonon coupled system is still missing,  
mainly due to the lack of proper description of spin-phonon and spin-photon couplings \cite{Metavitsiadis2021}.

Here, we make use of the $D_{3d}$ group symmetry of the Kitaev model \cite{You2012} and propose a theory to describe the Raman scattering of the Kitaev spin-phonon coupled system.
We show that our theory, in which 
the spin-phonon coupling and spin-photon coupling are explicitly built from the symmetry constraints,  quantitatively characterizes the temperature evolution and field dependence of the Fano lineshape of two low-energy optical phonons, observed in the Raman scattering experiments in $\alpha$-$\tx{RuCl}_3$ \cite{Sandilands2015,Glamazda2017, Li2019, Wulferding2020,Sahasrabudhe2020,Lin2020}.
These results reveal clear effects of the Majorana fermions and the $Z_2$ fluxes, which provide observable signatures for experimental detection of Kitaev QSL.

 \begin{figure}
 \includegraphics[width=\columnwidth]{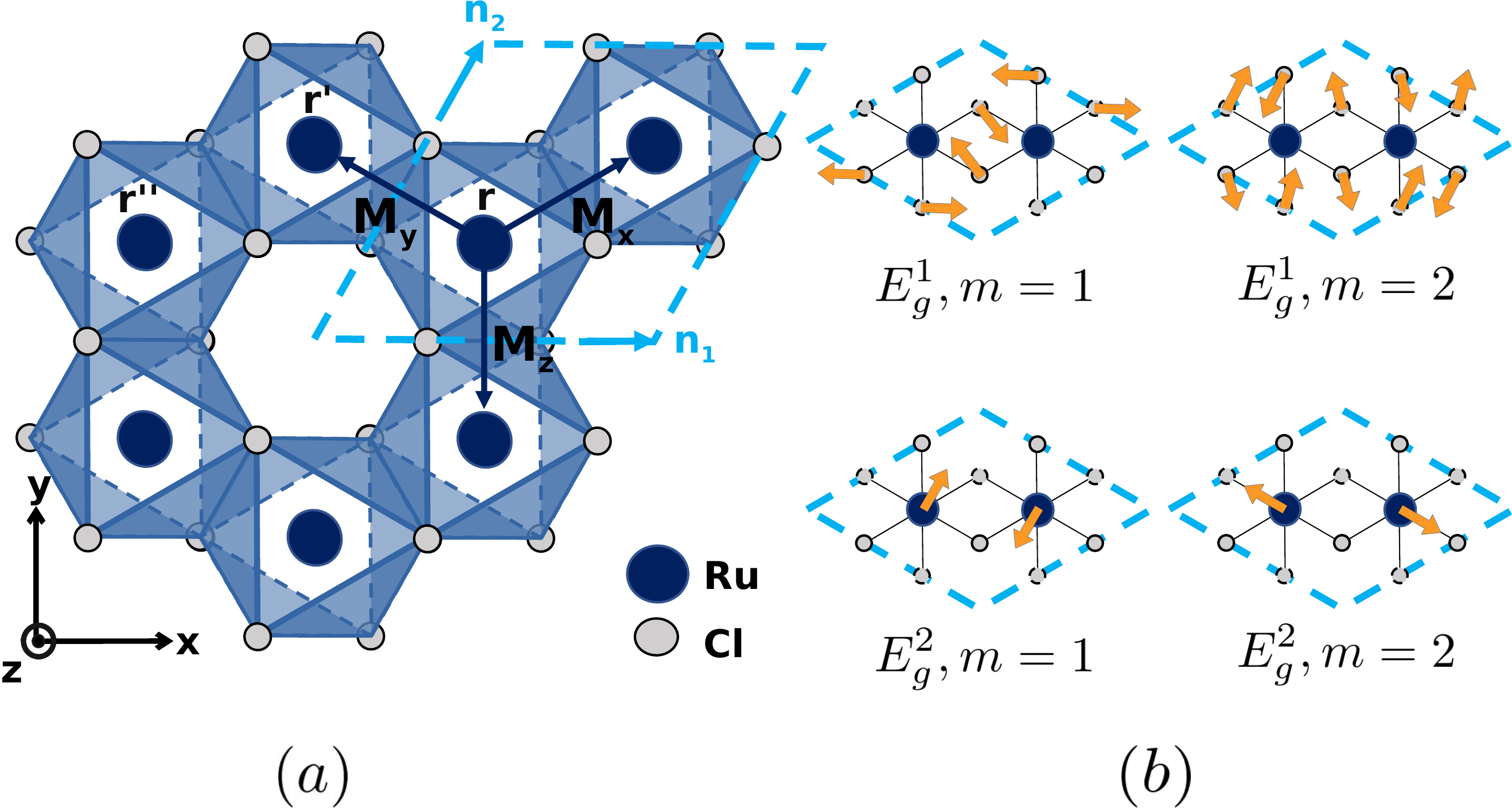}%
 \caption{
 (a) Crystal structure of $\alpha$-RuCl$_3$.
  The unit cell  shown in blue dashed lines is defined by  $\mathbf{n_1}=(\sqrt{3},0)$ and 
 $\mathbf{n_2}=\big(\frac{3}{2},\frac{\sqrt{3}}{2}\big)$ and includes two Ru$^{3+}$ and six Cl$^-$ ions.
 $\mathbf{M_{x,y}}=\big(\pm \frac{\sqrt{3}}{2},\frac{1}{2}\big)$ and $\mathbf{M_z}=(0,-1)$ are nearest neighbor vectors. The sites $\mathbf{r},\mathbf{r'},\mathbf{r''}$ form a generic three-spin link $\braket{\mathbf{r},\mathbf{r'},\mathbf{r''}}_{yx}$ as described in the text.
 (b) Visualization of the eigenmodes of $E^1_g$ and $E^2_g$ phonons in $xy$ plane, obtained by linear representation theory (see Sec.\,A of SM).
 }
 \label{Fig1}
 \end{figure}

%%%%%%%%%%%%%%%%%%%%%%%%%%%%%%%%%%%%%%%%%%%%%%%%%%%%%%%%
{\em Model.--}
We  consider the spin-phonon Hamiltonian
\begin{align}
\label{eq:model}
H=H_{\text s}+H_{\mathrm{ph}}+H_\tx{s-ph}.
\end{align}
The {\it first} term is the extended Kitaev honeycomb model \cite{Kitaev2006},
$H_{\text s}=$ $- J\sum_{\alpha,{\bf r}\in A}    \sigma_{\bf r}^{\alpha} \sigma_{{\bf r}+{\bf M}_\alpha}^{\alpha}- 
    \kappa \sum_{\langle {\bf r},{\bf r}',{\bf r}'' \rangle_{\alpha\gamma}}$ $   \sigma^\alpha_{{\bf r}} \sigma^\beta_{{\bf r}'} \sigma^\gamma_{{\bf r}''}$, 
  %\end{align}  
 where $\sigma^\alpha_{\bf r}$ are the Pauli matrices, $\alpha=x,y,z$ and ${\bf M}_\alpha$ are nearest neighbor vectors; $J$  denotes the Kitaev interaction; $\kappa$ is the strength of the time reversal symmetry breaking term, which mimics the effect of an external magnetic field \footnote{ 
% In our simulations we assume that the magnitude of the Kitaev interaction is $J=2meV\approx 23K$ as estimated in \cite{Sandilands2015}.
While we understand that the minimal model describing  describing $\alpha$-$\tx{RuCl}_3$ contains other terms \cite{Winter2017}, here we show that the main features  of the observed phonon dynamics can be understood already within the pure Kitaev model.}.
 The three-spin link notation $\langle {\bf r},{\bf r}',{\bf r}'' \rangle_{\alpha\gamma}$ 
 labels bonds ${\bf r}{\bf r}'$, ${\bf r}'{\bf r}''$  by type $\alpha,\gamma$ respectively and $\beta \neq \alpha, \gamma$. ${\bf r},{\bf r}',{\bf r}''$ are counter-clockwise ordered adjacent  sites (see \reffg{Fig1} (a)). 
 The leading order term in this Hamiltonian, i.e., the pure Kitaev model, has a symmetry described by $D_{3d}$ group 
 \footnote{The sixfold rotoreflection $S_6$ in $D_{3d}$, where $S_6 = C_{6}\sigma_h$ and $\sigma_h$ is a mirror reflection w.r.t the honeycomb plane, is a rotoreflection counterpart of the sixfold rotation $C_6$ in $C_{6v}$ group \cite{You2012}. See also the Supplementary Material \cite{Supplementarymaterial} for the explicit matrix forms of $D_{3d}$ group}.
%  The $\kappa$ term breaks the two-fold rotation around the in-plane axis and lowers the symmetry, but we still...
The $\kappa$ term lowers the symmetry to the $S_6$ group by breaking the two-fold rotation. But since we study both $\kappa=0$ and $\kappa > 0$ scienarios, we consider a $D_{3d}$-symmetric theory, which gives the strongest constraint. If the symmetry is broken into a subgroup of $D_{3d}$ \cite{koller2025spin}, a $D_{3d}$-symmetric theory would still be invariant under a lower-symmetry group.
 
$H_s$ is exactly solvable by the four Majorana fermion representation of spin \cite{Kitaev2006},  $\sigma_{\bf r}^{\alpha}=i  b_{\bf r}^{\alpha}c_{\bf r}$. In this representation,
$H_{\text s}=$ $
    \frac{1}{4} \sum_{\langle {\bf r}{\bf r}'\rangle} h_{{\bf r}{\bf r}'} c_{\bf r} c_{{\bf r}'}$,
where $ h_{{\bf r}{\bf r}'}=$ $  2 J\, i \eta_{{\bf r}{\bf r}'} +2 \kappa\, i\eta_{{\bf r}{\bf r}'} \eta_{{\bf r}'{\bf r}''} $ is the Hamiltonian matrix and $\eta_{{\bf r}{\bf r}'}=ib_{\bf r}^{\alpha} b_{{\bf r}'}^{\alpha} = \pm 1$ is the static $Z_2$ gauge field on the $\alpha$-bond, which generates  conserved $Z_2$ fluxes. Within each flux sector,
$H_\tx{s}$ can be further diagonalized to be
$H_{\tx s}=$ $ \sum_{i}\epsilon_i \left(\beta^\dagger_i \beta_i - 1/2\right)$, where $\epsilon_i$ are the fermionic energy levels and $\beta_i^\dagger,\beta_i$ correspond to the fermionic eigenmodes. 
Hereafter,
% \footnote{In our simulations we assume that the magnitude of the Kitaev interaction is $J=2meV\approx 23K$ as estimated in \cite{Sandilands2015}.
% While we understand that the minimal model describing  describing $\alpha$-$\tx{RuCl}_3$ contains other terms \cite{Winter2017}, here we show that the main features  of the observed phonon dynamics can be understood already within the pure Kitaev model.}
the energy and temperature unit will be $J$  unless  otherwise specified, which is estimated to be
$J\approx 2$ meV = 23 K
\cite{Sandilands2015}
\footnote{
 Note  that as  our model is written in terms of the Pauli matrices, the coupling constant $J$
 here is $1/4$ of the coupling for spin-1/2.}
.

The {\it second} term  in Eq.(\ref{eq:model}) is
the free phonon Hamiltonian
$H_{\mathrm{ph}}=$ $H_{\mathrm{ph}}\left(p_{i}(\vrr), q_{i}(\vrr)\right)$,
where $q_{i}( \vrr)= (x_1, y_1, z_1, \ldots, x_8, y_8, z_8)_{\bf r}$ denotes the displacement field in a unit cell at ${\bf r}$, which contains two Ru$^{3+}$ and six Cl$^-$ ions, shown in \reffg{Fig1}(a) and \reffg{unitcell} in the Supplementary Material (SM) \cite{Supplementarymaterial}; $p_{i}(\vrr)$ is the corresponding momentum.
Hereafter, we will drop the ${\bf r}$ dependence in phonon fields, since the long wavelength  of incident light leads to uniform lattice vibrations.
% By using the $D_{3d}$ symmetry of $\alpha$-RuCl$_3$,the eigenmodes of $H_\tx{ph}$  are written as linear superposition of the displacement fields in the irreducible representations (irreps) of this  group: $u_{\Gamma m} = \sum_{i=1}^{24} u_{\Gamma m, i} q_i$. Here, $\Gamma $ labels the irrep, i.e., 
By using the $D_{3d}$ symmetry of $\alpha$-RuCl$_3$, i.e.\ $[D_{3d}, H_\tx{ph}] = 0$,
the eigenmodes of $H_\tx{ph}$ are solved to be the irreducible representations (irreps) of the group, written as linear superpositions of the displacement field: $u_{\Gamma m} = \sum_{i=1}^{24} u_{\Gamma m, i} q_i$.
Here, $\Gamma $ labels the irrep, i.e.
$\Gamma=2 A_{1 g}+2 A_{2 g}+4 E_{g}+A_{1 u}+3 A_{2 u}+4 E_{u}
$, among which
the Raman active modes are $\Gamma_{R}=2 A_{1 g}+4 E_{g}$ \cite{Guizzetti1979,Li2019}, and $m$ is the dimension of the irrep. [See Sec.~A in the SM  for  detailed analysis \cite{Supplementarymaterial}].
%{\cbl
%By using the $D_{3d}$ symmetry of $\alpha$-RuCl$_3$:
% $[D_{3d}, H_\tx{ph}] = 0$,
%the eigenmodes of $H_\tx{ph}$ are solved to be the irreducible representations (irreps) of the symmetry group, written as linear %superpositions of the displacement field: $u_{\Gamma m} = \sum_{i=1}^{24} u_{\Gamma m, i} q_i$. Here, $\Gamma $ labels the irrep, %i.e.~$
%\Gamma=2 A_{1 g}+2 A_{2 g}+4 E_{g}+A_{1 u}+3 A_{2 u}+4 E_{u}
%$, among which
%the Raman active modes are $\Gamma_{R}=2 A_{1 g}+4 E_{g}$ \cite{Guizzetti1979,Li2019}. $m$ is the dimension of the irrep. See %Sec.~A in the SM.
%}
% In this work, we focus on the two low-energy phonon modes in the Raman spectroscopy \cite{Lin2020, Li2019, Sandilands2015}: $E_\tx{g}^1$ and $E_\tx{g}^2$, whose energies (around 14 meV and 20 meV respectively) are comparable to the magnetic continuum. These modes are both two-fold degenerate and are explicitly expressed as $u_{\Gamma m} = \sum_{i=1}^{24} u_{\Gamma m, i} q_i$, where $\Gamma = E_g^1, E_g^2$ denotes the irreps, and $m = 1, 2$ denotes degenerate modes (see Fig.\ref{Fig1} (b) for visualization).
% The corresponding free phonon Matsubara propagators can be written as
% {\it \cgr This paragraph break would cost two lines of space. It seems that not breaking the paragraph is ok to read.}
In this work, we focus on the two low-energy phonon modes in the Raman spectroscopy \cite{Lin2020, Li2019, Sandilands2015}:  $E_\tx{g}^1$ and $E_\tx{g}^2$, whose energies ($\sim$ 14 meV and 20 meV respectively) are comparable to the magnetic continuum's energy. 
% These modes are both two-fold degenerate and are visualized in Fig.\ref{Fig1} (b).
They are visualized in Fig.\ref{Fig1}(b).
The corresponding free phonon Matsubara propagators are written as
 $
     \mD^{(0)}_{\Gamma m, \Gamma'm'}(i\omega_n)=-\langle T_\tau u_{\Gamma m}(\tau)u_{\Gamma' m'}(0)\rangle_{\omega_n}  
     = \frac{2\omega_{\Gamma} }{\left(i \omega_{n}\right)^{2}-\omega_{\Gamma}^{2}  }   \delta_{\Gamma\Gamma'} \delta_{mm'}
 $, 
where $\omega_{\Gamma}$ is the frequency of the optical phonon, and $T_\tau $ is the imaginary time ordering operator.

%%%%%%%%%%%%%%%%%%%%%%%%%%%%%%%%%%%%%%%%%%%%%%%%%%%%%%%%%%%%

The {\it third} term  in Eq.(\ref{eq:model})  is the spin-phonon coupling Hamiltonian. It originates
 from the change of the Kitaev interaction in response to the lattice vibration: $J(q_i)=J + \sum_{\Gamma,m}\frac{\ud J(q_i)}{\ud u_{\Gamma m}} u_{\Gamma m}+\cdots$, where $\frac{\ud J(q_i)}{\ud u_{\Gamma m}}$ is the gradient along $u_{\Gamma m}$ direction 
 in the manifold of the displacement field.
The $D_{3d}$ invariant spin-phonon Hamiltonian is built as
\begin{align}
    H_\tx{s-ph}= \sum_{\Gamma, m} \lambda_\Gamma \Sigma_{\Gamma m} u_{\Gamma m}, \label{eq: Hsp}
\end{align}
where 
$\Sigma_{E_{g},1}=$ $\sum_\vrr (\sigma_{\mathbf{r}}^{x} \sigma_{\mathbf{r}+\mathbf{M}_{x}}^{x}+$ $\sigma_{\mathbf{r}}^{y} \sigma_{\mathbf{r}+\mathbf{M}_{y}}^{y}-$ $2 \sigma_{\mathbf{r}}^{z} \sigma_{\mathbf{r}+\mathbf{M}_{z}}^{z})$ and
    $\Sigma_{E_{g},2}=$ $ \sum_\vrr (-\sqrt{3}
        \sigma_{\mathbf{r}}^{x} \sigma_{\mathbf{r}+\mathbf{M}_{x}}^{x} 
        +$ $\sqrt{3}\sigma_{\mathbf{r}}^{y} \sigma_{\mathbf{r}+\mathbf{M}_{y}}^{y})$
are irreducible representations (irreps) of $D_{3d}$, and $\lambda_{\Gamma}$ are the coupling constants.
 
% {\em Renormalization of phonon propagator.--} 
As shown by the perturbative calculation in the SM,  the phonon propagator is renormalized by the spin-phonon coupling. According to  the Dyson's equation,  
${\hat \mD} =$ $ \leftr {({\hat \mD}^{(0)}})^{-1} -{\hat \Pi}  \rightr^{-1}$, where
${\hat \Pi}$ is the polarization bubble defined as 
\begin{align}
    \Pi_{\Gamma m, \Gamma' m'} &= -\lambda_\Gamma\lambda_{\Gamma' }\lefta T_\tau \Sigma_{\Gamma m} (\tau) \Sigma_{\Gamma' m'}(0) \righta . \label{eq: pol_bubble}
\end{align}
$\mD_{\Gamma m, \Gamma' m'}$ and $\Pi_{\Gamma m, \Gamma' m'}$ are 4 by 4 matrices, in which the $2\times2$ off-diagonal blocks correspond to the mixing between $E_g^1$ and $E_g^2$ phonon modes. 
The components of the  off-diagonal blocks are negligible, since the corresponding phonon peaks in the Raman spectroscopy are well separated \cite{Li2019}.

As will be seen later, the phonon  Raman peak parameters, such as the width, center position and asymmetry factor, are directly related to the real  and imaginary parts  of the fermionic loop diagrams contained in $\hat{\Pi}$ whose  temperature dependence at various values of $\kappa$ is shown in \reffg{unitcell} of SM. When temperature increases, both 
$\re \hat{\Pi}$ and  $\im \hat{\Pi}$,  evaluated at the bare phonon energies, generically display two-stage decrease  which is characterized by  two crossover temperatures. We can thus expect that this stage-wise temperature dependence in $\hat{\Pi}$ should be reflected in the temperature dependence of the phonon peak parameters, as shown next.

{\em Raman response.--} The Raman scattering of the  spin-phonon coupled Kitaev system (\ref{eq:model}) is described by the Raman operator: ${\mathcal R}$=$\sum_{\mu \mu'} \left({\mathcal R}^{\mu\mu'}_\tx{em-ph}+{\mathcal R}^{\mu\mu'}_\tx{em-s} \right)E^{\mu}_\tx{in} E^{\mu'}_\tx{out}$, where $E^\mu_\tx{in}$, $E^{\mu'}_\tx{out}$ are the electromagnetic fields of the incoming and outgoing light. The 
 second rank symmetric 
tensors ${\mathcal R}^{\mu\mu'}_\tx{em-ph}$ and ${\mathcal R}^{\mu\mu'}_\tx{em-s}$ microscopically describe the polarizability change of the electronic medium in response to the excitations of phonons and spins \cite{Dresselhaus2007}. Under the $D_{3d}$ symmetry constraint on the Raman operator, ${\mathcal R}^{\mu\mu'}_\tx{em-ph}$ is given by
\begin{align}
 {\mathcal R}^{\mu\mu'}_\tx{em-ph}=\sum_{ \Gamma, m} \mu_\Gamma  R_{\Gamma m}^{\mu \mu'} u_{\Gamma m}, \label{eq: HRemph}
\end{align}
where $R_{\Gamma m}$ are the Raman tensors taken from the irreps of $D_{3d}$, which are specified as
\begin{align}
    R_{E_g,1}^{\mu\mu'}=\left[\begin{array}{ccc}
c &0 &d \\
0& -c & 0\\
d& 0 &0
\end{array}\right], \ 
R_{E_g,2}^{\mu\mu'}=\left[\begin{array}{ccc}
0& -c &0  \\
-c &0  &d  \\
0&d  &0
\end{array}\right], \label{Ramantensor}
\end{align}
We take $c=1, d=0$ in the following computation. $\mu_\Gamma$ are the photon-phonon coupling constants. 
 The coupling of light to spins microscopically originates from its coupling to electric dipoles, which appears as a Wilson line operator that mediates the electronic hopping between the neighbouring ions \cite{Ko2010,Yang2021}.  Applying  the Loudon-Fleury approximation \cite{Fleury1968,Shastry1990}, %which is parallel to the derivation of the super-exchange interaction, 
 the magnetic part of the  Raman operator can be written as
\footnote{In a recent study \cite{Yang2021}, some of us showed that in the Kitaev candidate materials non-LF terms also appear in the  magnetic Raman scattering. However,  their main effects mainly appear at energies below $J$, so they will not change much physics at the energy scale above $J$. This is why here we constrain our consideration to the LF approximation.}
\begin{align}
 {\mathcal R}^{\mu\mu'}_{\tx{em-s}}=\nu\sum_{\alpha,{\bf r}\in A} {\bf M}_\alpha^{\mu}{\bf M}_\alpha^{\mu'} 
\sigma_{\bf r}^{\alpha} \sigma_{{\bf r}+{\bf M}_\alpha}^{\alpha},    \label{eq: HRems}
\end{align}
where $\nu$ is the photon-spin coupling constant.
${\mathcal R}^{\mu\mu'}_{\tx{em-s}}$  also satisfies the symmetry constraint, which can be  seen by decomposing it into the irreps of $D_{3d}$ as 
$ {\mathcal R}^{\mu\mu'}_{\tx{em-s}}$=$\nu \sum_{m}  R^{\mu\mu'}_{E_g,m} \Sigma_{E_g,m}$ (details in Sec.\,B of SM).

% where $\tilde{ R}^{\mu\mu'}_{E_g,m}$=$R^{\mu\mu'}_{E_g, m}(c=1,d=0)$.
% $ {\mathcal R}^{\mu\mu'}_{\tx{em-s}}$=$\nu \sum_{m}  \tilde{ R}^{\mu\mu'}_{E_g,m} \Sigma_{E_g,m}$, where $\tilde{ R}^{\mu\mu'}_{E_g,m}$=$R^{\mu\mu'}_{E_g, m}(c=1,d=0)$.

%%%%%%%%%%%%%%%%%%%%%%%%%%%%%%%%%%%%%%%%%%%%%%%%%%%%%%%%%%%%
In the {\it spin-phonon coupled system}, the Raman intensity is expressed in the interaction picture as $I(\Omega)\!=\!\int\!dt ~e^{i \Omega t} \langle T_t \mathcal{R}(t)\mathcal{R}(0) e^{-i\int \ud t' H_\tx{s-ph}(t')}\rangle$, where
$\langle \cdots\rangle\!=\!\text{Tr}[e^{-\beta{H_0}} \cdots ]/\text{Tr}[e^{-\beta{H_0}}]$ denotes the statistical average over 
the Hilbert space of the spin-phonon Hamiltonian $H_0 = H_s+H_\tx{ph}$,  $\beta\!=\!1/T$ is the inverse temperature, and $\Omega$ refers to the inelastic energy transfer by the photon.
Treating $H_\tx{s-ph}$ as perturbation, we perform systematic evaluation of the S-matrix expansion (see Sec.\,C of SM \cite{Supplementarymaterial}  for  explicit  derivations) and  obtain the  Matsubara Raman correlated function:
\begin{align}
    \mI (\tau)= \mI_\tx{em-s} (\tau) + R'_{L} (\tau) \cdot \hat{\mD} (\tau)\cdot R'_{ R} (\tau). \label{eq: Raman_inten}
\end{align}
Here, the dot product is on the contraction of ($\Gamma, m$) indices, ${R'}_{\Gamma m, L(R)}^{\mu\mu'}(\tau)$=$\mu_\Gamma R_{\Gamma m}^{\mu\mu'} +  \mP^{\mu\mu'}_{\Gamma m, L(R)}(\tau) $ are the  renormalized
 left and right phonon Raman vertices, which consist of
 the {\it bare}  phonon Raman vertex $ \mu_\Gamma R_{\Gamma m}^{\mu\mu'}$ and the {\it spin-dependent} phonon Raman vertex $\mP^{\mu\mu'}_{\Gamma m, L(R)}(\tau)$ \cite{Moriya1967, Suzuki1973}.
The bare phonon Raman vertex generates the phonon peak and constitutes the dominant contribution, while the spin-dependent phonon Raman vertex generates the salient Fano lineshape.
$\mI_\tx{em-s}^{\mu\mu'}(\tau)$=$-\langle T_{\tau} \mR_\tx{em-s}^{\mu\mu'}(\tau)\mR_\tx{em-s}^{\mu\mu'}(0)\rangle$ contributes to the magnetic continuum in the Raman spectrum. The physical Raman intensity is then obtained by the analytic continuation in the frequency domain: $i\Omega_n \to \Omega + i\delta_{\tx{ph}} $  followed by 
the application of the fluctuation-dissipation theorem.

% ============================================
% Fig.3
% ============================================
 \begin{figure}
 \includegraphics[width=1\columnwidth]{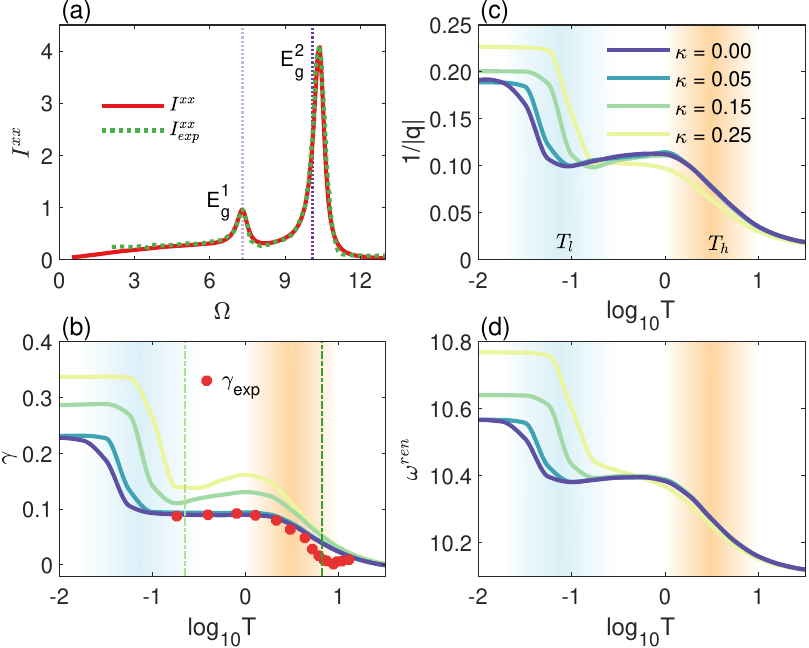}%
 \caption{
  Panel (a): $I^{xx}$  and $I^{xx}_\tx{exp}$ are, respectively,
 the stratified Monte Carlo (strMC) simulated  Raman intensity  and the experimental intensity from Ref.~\cite{Sandilands2015}
 at $T=0.22$ and $\kappa=0$. By fitting $I^{xx}$ to the experimental intensity $I^{xx}_\tx{exp}$, the best-fit model parameters are obtained: 
    $ \omega_\Gamma = [7.31, 10.10]$, 
    $\lambda_\Gamma = [0.25,    0.52]$,
    $\mu_\Gamma =   [0.38,    1.00]$, 
    $\nu = -0.63$.
Panels (b-d): The temperature dependence of the $E_g^2$ peak curve parameters obtained from the asymmetric Lorentzian fitting: $1/|q|$, 
$\gamma$ and $\omega^{\tx ren}$. 
$T_l$ and $T_h$ are two crossover temperatures. In panel (b), 
the computed $\gamma$ has been offset by a background line width obtained at $T=10^{1.5}$). This background line width mainly originates from the artificial broadening $\delta_\tx{ph}$ as shown in  Sec.\,D of SM.  The red dots are experimental line width $\gamma_\tx{exp}$, obtained from Ref.~\cite{Sandilands2015}.
The two green vertical dashed lines in (e) indicate $T=5$ K and 150 K. The unit conversion we use here is $J\approx 23$ K.
 }
\label{fig: FullModelFano}
\end{figure}

%%%%%%%%%%%%%%%%%%%%%%%%%%%%%%%%%
\begin{figure}
 \includegraphics[width=1\columnwidth]{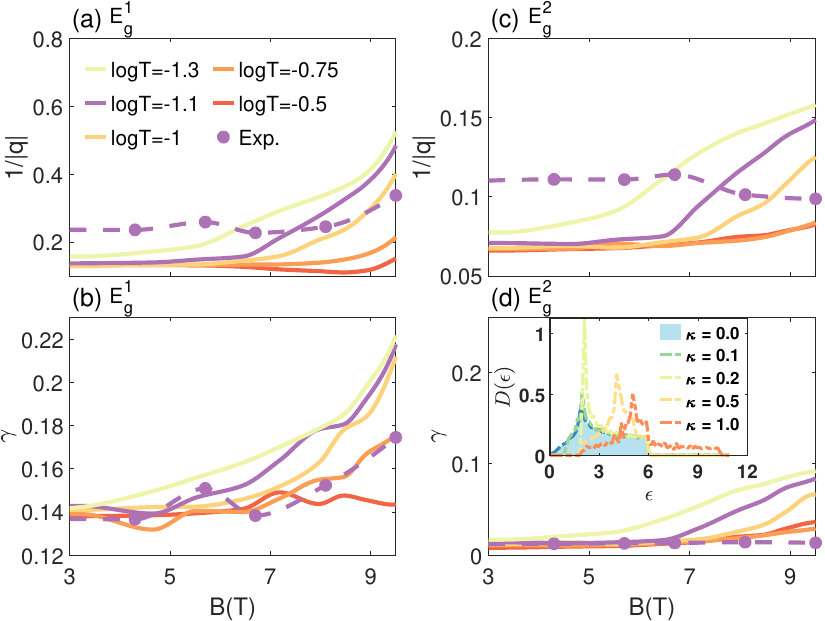}%
 \caption{
  The magnetic field dependence of curve parameters $1/|q|$ and $\gamma$ of two phonon peaks $E_g^1$ and $E_g^2$ in the computed Raman spectrum  are  shown in (a,b) and  (c,d), respectively. 
%   The curves are shown for various temperatures in the $T_l$ region. 
 The purple dots denote experimental data from Ref.~\cite{Wulferding2020} measured at $T=2$ K, i.e.~log$T=-1.1$. 
 The corresponding theoretical curve is also colored purple.
 The line width $\gamma$  has been offset by the background contribution (see caption of \reffg{fig: FullModelFano} for the reasoning). 
The inset of (b) shows the density of state of Majorana fermions at various $\kappa$ \cite{Feng2020}. The conversion from $\kappa$ in the unit of $J$ to magnetic field $B$ in the unit of Tesla %is the following. We assume [111] field direction for simplicity. Then
 follows from \cite{Kitaev2006}: $\kappa = \frac{(\mu_B B)^3}{\Delta_\tx{flux}^2}$, $\mu_B$ is the Bohr magneton and $\Delta_\tx{flux} = 0.27 J$ is the flux energy, and $J\approx 23$ K.
 }
 \label{Fig4_new}
 \end{figure} 
%%%%%%%%%%%%%%%%%%%%%%%%%%%%%%%%%%%%%%%%%%%%%%%%%%%%%%%%%%%
{\em Numerical results.---} 
With the developed formalism at hand, we now study the temperature evolution of the Raman spectrum and its $\kappa$ dependence with the focus on the Fano lineshape.
The thermodynamic average of the Raman correlation function  over different flux configurations is computed numerically
by using the stratified Monte Carlo (strMC) method \cite{Feng2021, feng2022phonon} on a lattice size of $N_1=N_2=25$. 
We will focus on  the $xx$-scattering geometry, in order to compare with the experiment, and assume $\delta_\tx{ph} = 0.15$.

To begin with, as shown in \reffg{fig: FullModelFano}(a), we first fit the computed Raman intensity $I^{xx}(\Omega)$ to the experimental Raman intensity $I^{xx}_\tx{exp}(\Omega)$ obtained from Ref.~\cite{Sandilands2015}, by tuning the adjustable {\it model parameters}   \{$ \omega_\Gamma , 
\lambda_\Gamma ,
\mu_\Gamma , 
\nu$\}, whose best-fit values are written in the caption of  \reffg{fig: FullModelFano}. $I^{xx}(\Omega)$ is obtained  by using \refeq{eq: Raman_inten} and evaluated at $T=0.22$ and $\kappa=0$.
%$I_\tx{exp}(\Omega)$ is obtained from Ref.~\cite{Sandilands2015} at the same temperature and $\kappa$.  
The details of the fitting procedure  and justification of
the uniqueness of the fitting parameters, after eliminating the overall scaling degree of freedom by setting $\mu_{E_g^2} = 1$, are described in Sec.\,E of SM. Remarkably, the best-fit parameter $\lambda_{E_g^2} = 0.52$ yields an estimation of the spin-phonon coupling to be $0.52 \times \!\sqrt{6} \!=\! 1.3J$, (with $\!\sqrt{6}$ being the norm of the bilinear $\Sigma_{\Gamma m}$ in \refeq{eq: Hsp}),  comparable to the first principle calculation of magnetoelastic coupling $\approx 4J$ given in Ref.~\cite{Kaib2021}. 

Next, with the fixed model parameters obtained above, we  compute  the evolution of the phonon Raman response by changing the temperature and  the strength of $\kappa$. 
%and the phonon Raman spectrum evolve accordingly. 
To quantitatively characterize the phonon peaks, we fit them to the asymmetric Lorentzian curve: 
$
I(\Omega)=
I_{0}\left[q \gamma+\left(\Omega-\omega^\tx{ren}\right)\right]^{2}/ [\gamma^{2}+\left(\Omega-\omega^\tx{ren}\right)^{2}]$,
where $1/q$ is the asymmetry factor, $\gamma$ 
is the half width at half maxima %(HWHM) 
which is referred to as line width hereafter, 
$\omega^{\tx ren}$ is the renormalized peak position and $I_0$ is the peak intensity.  

The temperature evolution of {\it curve parameters} $\{1/|q|, \gamma, \omega^{\tx ren}\}$ of the $E_g^2$ peak for various $\kappa$ is shown in \reffg{fig: FullModelFano} (b-d).  As mentioned above,  all curve parameters  display a  two-stage change with temperature. Two crossover temperatures, namely $T_l$ (in blue shaded area) and $T_h$ (in orange shaded area), correspond, respectively, to the flux proliferation temperature and the major fermionic excitation
temperature \cite{Feng2020, Nasu2015, Feng2021}.  In the $T_l$ region, the curve parameters $\{1/q, \gamma, \omega^{\tx ren}\}$ decrease significantly, which shows that they are sensitive to the emergent disorder from proliferated $Z_2$ fluxes. Also, the crossover temperature $T_l$ shows apparent $\kappa$ dependence, which reflects the increase of the flux gap energy with $\kappa$ \cite{Feng2020,Lahtinen2011}. In the $T_h$ region, further decrease of the curve parameters is due to Pauli exclusion principle of fermionic statistics.
  In \reffg{fig: FullModelFano}(b), we  also compare the experimental peak width $\gamma_{exp}$ obtained in Ref.~\cite{Sandilands2015} with the computed $\gamma$. Remarkably,  in the temperature region  between 5 K and 150 K we find  a good agreement  between them.
  This result indicates that the source of anomalous peak width observed in Ref.~\cite{Sandilands2015} can indeed be explained by spin-phonon coupling  within our theoretical framework. 
%   However, the phonon peak position $\omega^{ren}$ in Ref.~\cite{Sandilands2015} does not show the decrease from fermionic statistics. One possible explanation is that the renormalization of phonon peak position from spin-phonon coupling is already small at the lowest temperature in that experiment, so it will not show apparent further decrease as temperate increases.
Another noticeable result in \reffg{fig: FullModelFano}(b-d) is that at the lowest temperature, the curve parameters become larger with increasing $\kappa$. This is because, as magnetic field increases, more Majorana fermions become energetically comparable with the phonon modes (see the inset of \reffg{Fig4_new}(d)), and participate in the spin-phonon scattering. So the curve parameters become bigger.

%   {\cbl
%   Another noticeable result in \reffg{fig: FullModelFano}(b-d) is that at the lowest temperature, the curve parameters become larger with increasing field. This is because the magnetic field lifts the fermionic density of states $D(\epsilon)$ to higher energies, which can be seen in the inset of \reffg{Fig4_new}(d).
%   As a result, more Majorana fermions become energetically comparable with the two phonon modes located near the top of the fermion energy band, and participate in the spin-phonon scattering processes. So the curve parameters become bigger.

%   Since the energy of the two phonon modes are located near the high energy part of the continuum spectrum, as the field increases, more Majorana fermions become energetically comparable with the phonon modes and participate in the scattering processes.  So the curve parameters become bigger.
% }

The magnetic field dependence of $\{1/|q|, \gamma\}$ of $E_g^1$ and $E_g^2$ peak for various temperatures in the $T_l$ region is shown in \reffg{Fig4_new}. The conversion from $\kappa$ to external field $B$ is presented in the caption, where the field direction is assumed to be $[111]$ for simplicity. 
We can see a clear trend in both peaks that, for a larger temperature in the $T_l$ region, the curve parameters start to increase at a larger magnetic field. This is because $Z_2$ flux gap energy is proportional to $\kappa$; thus as the temperature  becomes larger, $Z_2$ fluxes require a higher magnetic field to be gapped out, after which the disorder introduced by $Z_2$ fluxes becomes  weaker and the Fano effects becomes stronger. So the curve parameters start to increase at a larger field. 
% We can see a clear trend that...
% It shows that the theoretical curve parameters generally increase with increasing magnetic field in the $T_l$ temperature region. This is because, for a fixed temperature, when the magnetic field becomes larger, the $Z_2$ fluxes are gapped out. So the disorder introduced by $Z_2$ fluxes becomes weak and the Fano effects become strong. Furthermore, there is a clear trend that, for a larger temperature in the $T_l$ region, the curve parameters start to increase at a larger magnetic field. 
% This is because $Z_2$ flux gap energy is proportional to $\kappa$ \cite{Kitaev2006}. So as temperature increases, they are gapped out at a higher field.
%{\cbl
%{\it \cgr It seems that the paragraph would be too heavy to read without this break.}  

The computed curve parameters can be compared with the low-temperature experimental from Ref.\cite{Wulferding2020} and Ref.~\cite{Sahasrabudhe2020}. 
The data from Ref.\cite{Wulferding2020} is shown in 
\reffg{Fig4_new}
in the magnetic field region $B=3\sim 9T$ containing the putative QSL phase. 
Remarkably, in \reffg{Fig4_new}(a-b) there is a discernible increase in the parameters $\{1/|q|, \gamma\}$ in $E_g^1$ peak, whose magnitude is comparable with the theoretical increase. 
Our results also suggest that if the increase of the curve parameters at higher temperatures starts at  higher fields, then
 this observation is consistent with the behaviour of the $Z_2$ fluxes.
%{\cbl
%Furthermore, if curve parameters of the $E_g^1$ peak are measured for multiple temperatures in the $T_l$ region, then the trend %that gapped-out magnetic field increases with temperature could be expected, which would would a consistent with the behaviour of %the $Z_2$ fluxes.
%}
%This indicates that the field dependence of the curve parameters is experimentally probable at least in $E_g^1$ peak.
%}
%\reffg{Fig4_new}(c-d) shows the field dependence of the $E_g^2$ peak curve parameters.  While  the computed curve parameters show the behavior similar  to the $E_g^1$ peak parameters, {\it \cgr This is somewhat repetitive to the contents in the paragraph above}
%{\cbl
In \reffg{Fig4_new}(c-d), the experimental field dependence of the $E_g^2$ peak curve parameters remains featureless. This could be attributed to the fact that the $E_g^2$ phonon has higher energy than $E_g^1$, thus it is less sensitive to the increased population of fermionic modes from the increased field. 
%}
%the experimental field dependence of the $E_g^2$ peak  parameters remain featureless.

%we can propose experiments that measure the field dependence of the curve parameters for more temperatures in the $T_l$ region. If the trend that, for a larger temperature in the $T_l$ region, the curve parameters start to increase at a larger magnetic field were confirmed, then this result would be consistent with the behaviour of the $Z_2$ fluxes.

%{\cbl
%Furthermore, we can propose experiments that measure the field dependence of the curve parameters for more temperatures in the %$T_l$ region. If the trend that, for a larger temperature in the $T_l$ region, the curve parameters start to increase at a larger %magnetic field were confirmed, then this result would be consistent with the behaviour of the $Z_2$ fluxes.
%}

 {\it Conclusion. --}
 We have constructed a theory % with $D_{3d}$ symmetry 
 to describe the Raman scattering of the spin-phonon coupled Kitaev system. Based on this theory, we systematically compute the Raman spectrum and explore the temperature evolution and the magnetic field dependence of the phonon peaks in Raman spectrum, which are consistent with the Raman scattering experiment in $\alpha\tx{-RuCl}_3$. Our theory clarifies the mechanism of how spin-phonon coupling generates Fano lineshapes, and also offers an estimate of the spin-phonon coupling by model fitting. These results open the possibility of experimentally identifying the effects of fractionalized excitations of QSL hidden in the Fano lineshapes of phonon Raman peaks.

%============================================
% \section{Fano effect}

%%%%%%%%%%%%%%%%%%%%%%%%%%%%%%%%%%%%%%%%%%%%%%%%%%%%%%%%%

{\it  Acknowledgments:}  The authors are thankful to
Ken Burch, Jia-Wei Mei, Joji Nasu,  Kenya Ohgushi, Thuc T Mai, Luke Sandilands, Yiping Wang, Yang Yang, Mengxing Ye, Shuo Zhang  and especially Dirk Wulferding for valuable discussions.
The work was  supported by  the  U.S. Department  of  Energy,  Office  of  Basic  Energy Sciences  under  Award  No. DE-SC0018056. 
   N.B.P. acknowledges the hospitality of Aspen Center of Physics.
%%%%%%%%%%%%%%%%%%%%%%%%%%%%%%%%%%%%%%%%%%%%%%%%%%%%%%%%%
 \bibliography{bib} 
% \end{document}
%%%%%%%%%%%%%%%%%%%%%%%%%%%%%%%%%%%%%%%%%%%%%%%%%%%%%%%%%
%%%%%%%%%%%%%%%%%%%%%%%%%%%%%%%%%%%%%%%%%%%%%%%%%%%%%%%%%
\clearpage
\onecolumngrid

\appendix

\begin{center}
{ \bf \Large Supplementary Material} 
\end{center}

\setcounter{figure}{0}
% \numberwithin{equation}{section}
\renewcommand{\theequation}{\thesection\arabic{equation}}
\renewcommand{\thefigure}{S\arabic{figure}}

\setcounter{equation}{0}
\renewcommand{\thesection}{A}
\section{A. The irreducible representations of the phonon modes}\label{App:1}

In the main text, we have introduced the phonon Hamiltonian $H_\tx{ph}$. Its normal vibration modes will be solved by group theory. The point group we consider here is $D_{3d}$, which is the symmetry shared by both the Kitaev model \cite{You2012} and a single layer of $\alpha$-$\tx{RuCl}_3$ \cite{Li2019}. The invariance of the phonon Hamiltonian under the group operations requires $[H_{\mathrm{ph}}, D_{3d}]=0$. 
Then, to obtain the eigenmodes of $H_\tx{ph}$, we apply the following theorem:
\begin{myTheo}
    If a Hamiltonian $H$ is invariant under the group $G$, i.e., $[G,H] = 0$, then the irreducible representation of $G$ forms the basis of the eigensubspace of $H$; and the energy of  multidimensional irreducible representation is degenerate.
\end{myTheo}
The proof 
can be found in Ref.\ \cite{feng2022phonon, Inui2012}
\footnote{
Equivalent to Schur's Lemma in group representation theory.
}.
% Careful readers will notice that the proof in Ref.\ \cite{Inui2012} is not directly on the theorem here but rather on its converse version, i.e.\ {\it the eigenstates of the Hamiltonian form a basis for an irreducible representation of the symmetry group}. This proof should suffice to illustrate the physics. But we do have a strict proof of the theorem, which is too technical to put here. It is available on request. 
% }.
Applying this theorem to the current work, we can see that  in the symmetry group $D_{3d}$,
 the irreducible representation of  normal vibration modes $u_{\Gamma m}$ forms the eigensubspace of $H$,
and the energy of 2-dimensional irreducible representation $E_g$ is degenerate, i.e., $H_\tx{ph} u_{\Gamma m} = \omega_{\Gamma} u_{\Gamma m}$, where $\Gamma= E_g$.

\begin{figure}
 \includegraphics[width=0.4\columnwidth]{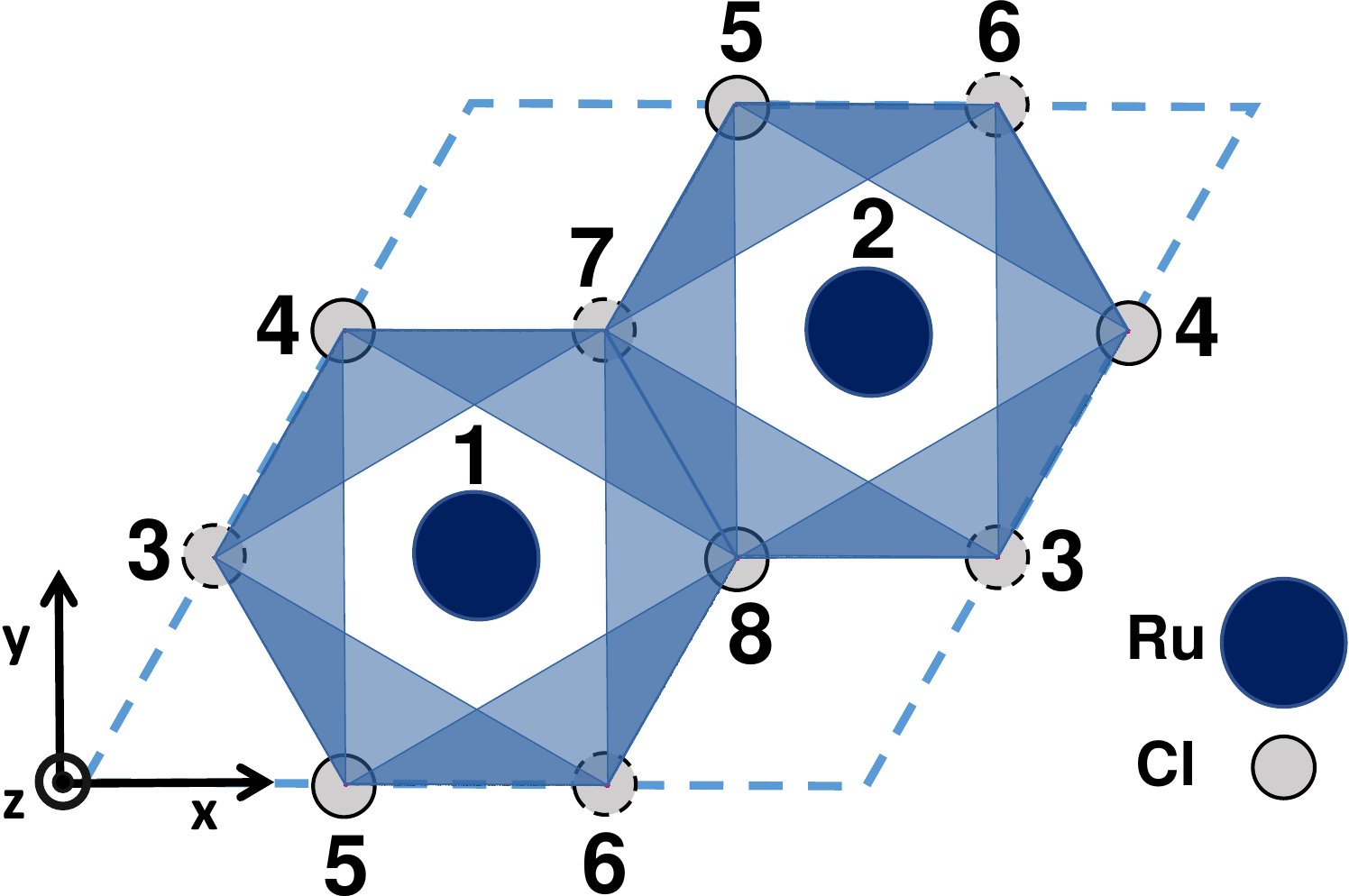}
 \caption{Unit cell of $\alpha-$RuCl$_3$ with labeled ions.}
 \label{unitcell}
 \end{figure}

As introduced in the main text, the general form of the phonon Hamiltonian can be written as $H_\tx{ph} = H_\tx{ph}(p_i(\vrr), q_i(\vrr))$, where $q_{i}( \vrr)= (x_1, y_1, z_1, \ldots, x_8, y_8, z_8)_{\bf r}$ describes the displacement fields in a unit cell located at $\vrr$, which contains two Ru$^{3+}$  and six Cl$^{-}$ ions shown in \reffg{unitcell}. $p_{i}(\vrr)$ is the corresponding momentum.
The vibration eigenmodes at the center of the Brillouin zone is classified according to the irreducible representations (irreps) of $D_{3d}$:
$
    \Gamma=2 A_{1 g}+2 A_{2 g}+4 E_{g}+A_{1 u}+3 A_{2 u}+4 E_{u}
$, among which
the Raman active modes are $\Gamma_{R}=2 A_{1 g}+4 E_{g}$ \cite{Guizzetti1979,Li2019}. 
%Note that this classification should in principle apply for the most general form the phonon Hamiltonian $H_p(p_i, q_i)$ with the symmetry constraint. 
Assuming that the phonon potential energy  can be expanded as $V(q_i) = 
V_0 + \onefrac{2}\sum_{i,j=1}^{24}\left.\frac{\partial^2 V}{\partial { q}_i\partial { q}_j}\right|_0 {q}_i { q}_j + \cdots$,  then  a  vibration eigenmode
can be written as a linear combination of the displacement fields: $u_{\Gamma m}(\vrr) = \sum_i u_{\Gamma m, i} { q}_i(\vrr)$, where
$\Gamma$ denotes the irreps of dimension $m$. As mentioned in the main text, we will drop the $\vrr$ dependence due to long wave approximation. Then, applying linear representation theory of finite groups \cite{Inui2012, Dummit2004, Serre1977} \footnote{The programming code for this computation is available upon request.}, we obtain the explicit form of these vibration modes and show the Raman active ones here:
% are  interested in the phonon Raman response originated from  the vibration modes  with ${\bf q}=0$,  similarly to Refs.\cite{Hasegawa2017,Mai2019}
%  we can explicitly obtain  them by   using the linear representation theory of finite groups \cite{Inui2012, Serre1977}:
\begin{align}
&E^1_{g}:\left\{
   \begin{array}{l}
     0.07 x_3 -0.47 z_3 -0.07 x_4 +0.47 z_4 -0.27 x_5 -0.11 y_5 -0.23 z_5 +0.27 x_6    \\ \quad
     +0.11 y_6 +0.23 z_6+0.27 x_7 -0.11 y_7 +0.23 z_7 -0.27 x_8 +0.11 y_8 -0.23 z_8  ,
     \\
     0.33 y_3 -0.33 y_4 -0.11 x_5 -0.14 y_5 +0.40 z_5 +0.11 x_6 \\ \quad
    +0.14 y_6 -0.40 z_6-0.11 x_7 +0.14 y_7 +0.40 z_7 +0.11 x_8 -0.14 y_8 -0.40 z_8 ,
\end{array}
\right.\label{eq: Eg1}
\\
&E^2_{g}:\left\{
\begin{array}{l}
         \onefrac{\sqrt{2}} (y_1 - y_2), \\
         \onefrac{\sqrt{2}}(- x_1 + x_2),
    \end{array}
\right.\label{eq: Eg2}
\\
&E^3_{g}:\left\{
    \begin{array}{l}
 -0.57 x_3 -0.06 z_3 +0.57 x_4 +0.06 z_4 +0.11 x_5 -0.27 y_5 -0.03 z_5 -0.11 x_6 \\ \quad
 +0.27 y_6 +0.03 z_6 -0.11 x_7 -0.27 y_7 +0.03 z_7 +0.11 x_8 +0.27 y_8 -0.03 z_8  ,
 \\
 0.04 y_3 -0.04 y_4 -0.27 x_5 +0.42 y_5 +0.05 z_5 +0.27 x_6 \\ \quad
 -0.42 y_6 -0.05 z_6 -0.27 x_7 -0.42 y_7 +0.05 z_7 +0.27 x_8 +0.42 y_8 -0.05 z_8 ,
    \end{array}
\right.\label{eq: Eg3}
\\
&E^4_{g}:\left\{
\begin{array}{l}
     -0.34 z_3 +0.34 z_4 +0.35 x_5 +0.20 y_5 -0.17 z_5 -0.35 x_6   \\ \quad
     -0.20 y_6 +0.17 z_6 -0.35 x_7 +0.20 y_7 +0.17 z_7 +0.35 x_8 -0.20 y_8 -0.17 z_8  ,
     \\
      -0.47 y_3 +0.47 y_4 +0.20 x_5 +0.12 y_5 +0.29 z_5 -0.20 x_6\\ \quad
    -0.12 y_6 -0.29 z_6 +0.20 x_7 -0.12 y_7 +0.29 z_7 -0.20 x_8 +0.12 y_8 -0.29 z_8,
\end{array}
\right.\label{eq: Eg4}
\\
&A^1_{1g}:
    \begin{array}{l}
 -\frac{1}{\sqrt{2}} z_1 +\frac{1}{\sqrt{2}} z_2 ,
    \end{array} \label{eq: A1g1}
\\
&A^2_{1g}:
 -\onefrac{\sqrt{6}}x_3 +\onefrac{\sqrt{6}}x_4 -\onefrac{2\sqrt{6}} x_5 +\onefrac{2\sqrt{2}} y_5 +\onefrac{2\sqrt{6}} x_6 -\onefrac{2\sqrt{2}} y_6 +\onefrac{2\sqrt{6}} x_7 +\onefrac{2\sqrt{2}} y_7\\ \nn
 & \qquad\qquad
 -\onefrac{2\sqrt{6}} x_8 -\onefrac{2\sqrt{2}} y_8 , \label{eq: A2g1}
\end{align}
where the atom labeling convention followed is shown in \reffg{unitcell}. The explicit solution for infrared-active modes were obtained in a similar way in Ref.\ \cite{Hasegawa2017}. 

To make sense of the phonon modes solutions shown above, we consider a concrete example of phonon Hamiltonian
$H_\tx{ph} = \sum_i \frac{p_i^2}{2m_i} +  V(q_i)$, where $m_i$ is the mass of the i-th vibrating coordinate, and the potential energy is quadratic as introduced above: $V(q_i) =  \onefrac{2}\sum_{i,j=1}^{24}\left.\frac{\partial^2 V}{\partial { q}_i\partial { q}_j}\right|_0 {q}_i { q}_j $.
Then
$H_\tx{ph}$ is rewritten as: 
\begin{align}
    H_\tx{ph} =  \sum_{ij} \left[ \onefrac{2} \tilde{p}_i\one_{ij}\tilde{p}_j + \onefrac{2} \tilde{q}_iK_{ij}\tilde{q}_j \right],
\end{align} 
where $\tilde{p}_i = \frac{p_i}{\sqrt{m_i}}$, $\tilde{q}_i = \sqrt{m_i}q_i$ are conjugate canonical coordinates, and $K_{ij} = \left.\frac{\partial^2 V}{\partial { q}_i\partial { q}_j}\right|_0 \cdot \onefrac{\sqrt{m_i m_j}}$. So, if this phonon Hamiltonian $H_\tx{ph}$ satisfies the $D_{3d}$ symmetry (mainly the potential energy term, since the kinetic energy term is isotropic), then the quadratic potential energy $K_{ij}$ will be block-diagonalized by the modes $u_{\Gamma m}$ listed above, with each block being labelled by the irreducible representation $\Gamma$.
It is then clear that the modes $u_{\Gamma m}$ indeed give the correct eigen vibration modes for the phonon Hamiltonian $H_\tx{ph}$. The following steps are generic: $H_\tx{ph} = \sum_{\Gamma m}\onefrac{2}\omega_\Gamma \left[\frac{P_{\Gamma m}^2}{\omega_\Gamma} + \omega_\Gamma u^2_{\Gamma m} \right] = \sum_{\Gamma m} \omega_\Gamma (b_{\Gamma m}^\dagger b_{\Gamma m} + \onefrac{2})$, where $\omega^2_{\Gamma}$ is the eigenvalue of $K$, $P_{\Gamma m} = \sum_i u_{\Gamma m, i} \tilde{p}_i$ and $u_{\Gamma m} = \sum_i u_{\Gamma m, i}\tilde{q}_i$ are conjugate canonical coordinates of the eigenmodes, and $b_{\Gamma m}^\dagger, b_{\Gamma m}$ are the corresponding creation and annihilation operators.

% In the basis of these modes, a quadratic form of phonon potential energy $V(q_i) =  \onefrac{2}\sum_{i,j=1}^{24}\left.\frac{\partial^2 V}{\partial { q}_i\partial { q}_j}\right|_0 {q}_i { q}_j $ that satisfies $D_{3d}$ symmetry will be block-diagonalized, with each block being labelled by irreducible representation $\Gamma$. 
% Thus, the vibration eigenmodes have been obtained. Then the phonon Hamiltonian can be written as: $H_\tx{ph} = \sum_{\Gamma m} \omega_{\Gamma}(b_{\Gamma m}^\dagger b_{\Gamma m} + \onefrac{2})$, where $\omega_{\Gamma}$ denotes the eigen energy of the $\Gamma$ phonon and $b_{\Gamma m}^\dagger, b_{\Gamma m}$ are the creation and annihilation operators for modes $u_{\Gamma m}$.

% Note that, if we consider the phonon sector alone, then each $E_g$ irrep
% %the above basis of the irreducible representations each
% has freedom of rotation by $O(2)$ matrices without changing the block-diagonal structure of the representation matrix. However, when we consider the $E_g$ irrep of all three sectors together, namely, the phonons, the spin bilinear products and the Raman polarization tensors,
% the basis of each sector needs to be fixed, such that the representation matrix are the same in all three sectors. Thus, the coupling Hamiltonians built by the inner product of two sectors are invariant under group operation.
% % $\vec{e}\cdot \vec{b} \to \vec{e} O \cdot O^\T \vec{b}$
% In this sense, the form of the irreps  (\ref{eq: Eg1}-\ref{eq: A2g1}) is  fixed, unless the basis of all three sectors are rotated by $O(2)$ simultaneously.

 %Assuming the quadratic form of the potential energy above is a valid approximation, 
Next we compare the above phonon modes with the results from the density functional theory (DFT) calculations \cite{Li2019}
\footnote{The explicit result is shared by the author through private communication.} with a goal to identify the two low-energy $E_g$ modes among
 %First, there are indeed 
the four pairs of $E_g$ modes identified by DFT. By looking at the major dominant vibrating components and their relative directions 
on each Ru$^{3+}$ and Cl$^{-}$ ions, we conclude 
% By comparing the energy degeneracy, energy ordering and features of the vibration, we identified 
that the low-energy modes $E_g^1$ and $E_g^2$ modes are those given by
%considered in \cite{Li2019} {\cbl are the closest to}
\refeq{eq: Eg1} and \refeq{eq: Eg2}, respectively.
Note that the $E^2_g$ mode \refeq{eq: Eg2} only involves vibrations of $\tx{Ru}^{3+}$ ions. 
%Note that, 
If only the vibration of $\tx{Ru}^{3+}$ ions are considered \cite{Metavitsiadis2021},  under $D_{3d}$ constraint  the phonon modes 
decompose as $\Gamma_{Ru} = A_{1g} + E_{g} + A_{2u} + E_{u}$, where $A_{2u} + E_{u}$ are the acoustic modes, and $E_{g}=E_g^2$ and $A_{1g}$ modes are those given by \refeq{eq: Eg2} and \refeq{eq: A1g1}.
% Contrary to intuition, the DFT calculations \cite{Li2019} suggest that this mode has higher energy than $E^1_g$ mode \refeq{eq: Eg1}, in which  vibrations are predominantly from $\tx{Cl}^{-1}$ ions.
The DFT calculations \cite{Li2019} suggest that $E_g^2$ has higher energy than $E^1_g$ mode \refeq{eq: Eg1}, in which vibrations are predominantly from $\tx{Cl}^{-1}$ ions whose mass is smaller. This indicates that the corresponding stiffness is also smaller.

Finally, we give the matrix representations of the $D_{3d}$ group in the basis of the $E_g$ phonons Eq.(\ref{eq: Eg1})-(\ref{eq: Eg4}), i.e.\ $\bra{u_{E_g, m}} D_{3d} \ket{u_{E_g, m'}}$, where $ m,m'=1$, 2. It suffices to just show the result of the two generators of $D_{3d}$: $S_6$ the 6-fold rotoreflection, and $C'_2$ the 2-fold rotation around the $y$-axis in \reffg{unitcell}, while the representations of other group elements can be obtained via group multiplication. In the basis of $E_g$ phonons Eq.(\ref{eq: Eg1})-(\ref{eq: Eg4}), we obtain the following matrix representation,
\begin{align}
S_6=\left[\begin{array}{cc}
-1/2        &\sqrt{3}/2 \\
-\sqrt{3}/2 & -1/2 
\end{array}\right],\quad   
C'_2 =\left[\begin{array}{cc}
1& 0  \\
0 &-1  
\end{array}\right]. \label{Eg_mat}
\end{align}
This $E_g$ representation in basis of the phonon sector is exactly the same as the $E_g$ representation in the basis of the other two sectors, namely the spin bilinear products $\Sigma_{\Gamma m}$ explicitly written under \refeq{eq: Hsp}, and the Raman polarization tensors $R_{\Gamma m}^{\mu\nu}E_\tx{in}^{\mu}E_\tx{out}^{\nu}$ introduced in \refeq{Ramantensor} with $c=1, d=0$. This guarantees that the coupling Hamiltonians built by the inner product of the basis from any two of the aforementioned three sectors are invariant under the $D_{3d}$ transformations.

The explicit matrix representation in basis of the above three sectors is obtained in the following computation. First, for a given element $g\in D_{3d}$, the transformation of the physical fields involved in the three sectors is specified by the following expressions,
\begin{align}
    \vec{\sigma}(\vrr) &\to \vec{\sigma}\,'(\vrr') =  O_g^\T\vec{\sigma}(O_g\vrr), \\
    \vec{E}(\vrr) &\to \vec{E}\,'(\vrr') = O_g^\T \vec{E}(O_g\vrr),\\
    \vec{q}\,(\vrr) &\to \vec{q}\,'(\vrr') = O_g^\T\vec{q}\,(O_g\vrr).
\end{align}
Here, $O_g$ is an operation of $g$ on a 3D vector, including coordinate $\vrr$, $\vec{\sigma}$ and $\vec{E}$, and can be represented by a $3\times3$ matrix. When $O_g$ applies on $\vec{q}$, a $24$-dimensional vector, $O_g$ is represented as a $24\times24$ matrix, which is an operation on the whole unit cell containing 6 ions, each with 3D vibration degrees of freedom. Then, the transformation of basis from the spin bilinear products $\Sigma_{\Gamma m}(\vrr) $, the Raman polarization tensors $R_{\Gamma m}^{\mu \nu} E^{\mu}_\tx{in}(\vrr) E_\tx{out}^{\nu}(\vrr)$, and the phonon sector $u_{\Gamma m, i}q_i(\vrr)$ can be derived, and the corresponding matrix representation of $D_{3d}$ group under these basis can be obtained. Based on this, the $D_{3d}$ invariance of the coupling Hamiltonians \refeq{eq: Hsp} \refeq{eq: HRemph} \refeq{eq: HRems}, and the $E_g$ representation \refeq{Eg_mat} can be explicitly verified.

\setcounter{equation}{0}
\renewcommand{\thesection}{B}
\section{B. The symmetry decomposition of the Loudon-Fleury Raman operator}

As introduced in the main text, the Loudon-Fleury Raman operator has $D_{3d}$ symmetry. To explicitly see this symmetry, here we show the explicit decomposition of this operator into the irreducible representations of $D_{3d}$. 

To begin with, we first introduce the irreducible representations of the Raman tensors \cite{kroumova2003bilbao}\footnote{Note that, here 
% it appears differently from Ref.\ \cite{djurdjic2018lattice} since 
the coordinate system as illustrated in the figure has been rotated from the standard settings of space group.}:
\begin{align}
R_{A_{1g}}^{\mu\mu'}=\left[\begin{array}{ccc}
a &0 &0 \\
0& a &0 \\
0& 0& b
\end{array}\right] ,  \quad 
R_{E_g,1}^{\mu\mu'}=\left[\begin{array}{ccc}
c &0 &d \\
0& -c & 0\\
d& 0 &0
\end{array}\right], \quad  
R_{E_g,2}^{\mu\mu'}=\left[\begin{array}{ccc}
0& -c &0  \\
-c &0  &d  \\
0&d  &0  
\end{array}\right]. \label{Ramantensor_all}
\end{align}
In the following derivative, we use $a=c=1, b=d=0$ since we consider only 2D component of electromagnetic field. 

Then as introduced in the main text, the Raman operator of this spin-phonon coupled Kitaev system is described by ${\mathcal R}$=$\sum_{\mu \mu'} \left({\mathcal R}^{\mu\mu'}_\tx{em-ph}+{\mathcal R}^{\mu\mu'}_\tx{em-s} \right)E^{\mu}_\tx{in} E^{\mu'}_\tx{out}$, and the coupling of electromagnetic wave to spin is described by Loudon-Fleury operator
$
 {\mathcal R}^{\mu\mu'}_{\tx{em-s}}=\nu\sum_{\alpha,{\bf r}\in A} {\bf M}_\alpha^{\mu}{\bf M}_\alpha^{\mu'} 
\sigma_{\bf r}^{\alpha} \sigma_{{\bf r}+{\bf M}_\alpha}^{\alpha}
$.
Now we are ready to decompose this tensor into the irreducible representations in \refeq{Ramantensor_all}: $\mR^{\mu\mu'}_{\tx{em-s}} = \sum_{\Gamma m} \alpha_{\Gamma m} R^{\mu\mu'}_{\Gamma m}$, where $R_{\Gamma m}^{\mu\nu}$ are the Raman tensors. Using their orthogonality relations, the coefficient $\alpha_{\Gamma m}$ are obtained: $\alpha_{\Gamma m} = \onefrac{2}\Tr[R_{\Gamma m}^{\ \T} \cdot \mR_\tx{em-s}]$, where the dot is the matrix product on $\mu\mu'$ indices.
Then the symmetry decomposition of  ${\mathcal R}_{\tx{em-s}}$  according to $D_{3d}$ can be written as
$\mR^{\mu\mu'}_{\tx{em-s}} = \nu \sum_{\Gamma m} \Sigma_{\Gamma m} R^{\mu\mu'}_{\Gamma m}$.
Since $\left[{\mathcal R}_{\tx{em-s}, A_{g}},H_{\tx s}\right]=0$, and thus  only $\Gamma=E^1_{g}$ and $\Gamma=E^2_{g}$ channels contribute into the Raman response with the Raman operator given by ${\mathcal R}_{\tx{em-s}}^{\mu\mu'} =\nu \sum_{ m} \Sigma_{E_g, m}  R^{\mu\mu'}_{E_g, m}$.

\setcounter{equation}{0}
\renewcommand{\thesection}{C}
\section{C. Perturbative calculation of the Raman response in the spin-phonon coupled  Kitaev system} 
 
The  spin-dependent phonon  Raman scattering intensity is calculated as follows. 
There are two channels for the Raman scattering response, the phonon and the spin, so the Raman operator can be written as
\begin{align}
  {\mathcal R}=  {\mathcal R}_\tx{em-ph}+ {\mathcal R}_\tx{em-s},
\end{align}
where  ${\mathcal R}_\tx{em-ph} = \sum_{ \Gamma, m} \mu_\Gamma  R_{\Gamma m}^{\mu \mu'} u_{\Gamma m}E^{\mu}_\tx{in} E^{\mu'}_\tx{out}$ and ${\mathcal R}_\tx{em-s} = \nu \sum_{m}   R^{\mu\mu'}_{E_g,m} \Sigma_{E_g,m}E^{\mu}_\tx{in} E^{\mu'}_\tx{out}$, respectively, denote the coupling of 
electromagnetic field of light to phonons and spins as introduced in the main text. % Here $\tilde{ R}^{\mu\mu'}_{E_g,m}$ is also defined in Eq.(B5)

In the simplest case, where the spins and phonons are decoupled, the Raman intensity is expressed as 
 \begin{align}
 I(\Omega)\!=\!\int\!dt ~e^{i \Omega t} \langle T_t\mathcal{R}(t)\mathcal{R}(0) \rangle,
 \end{align}
 where 
$ I(t) =  \left\langle T_t \mR(t) \mR(0)  \right\rangle$ is the time-ordered Raman correlation function.
 It is also convenient to introduce  the retarded Raman correlation function \cite{Mahan}, which is also known as the Raman susceptibility:
\begin{align}\chi(\Omega) = -i\int \ud t e^{i\Omega t} \Theta(t)\langle \leftr \mR(t), \mR(0) \rightr\rangle. 
 \end{align}
 The Raman intensity $I(\Omega)$  and the Raman susceptibility  $\chi(\Omega)$ are related via the fluctuation-dissipation theorem: 
\begin{align}
    I(\Omega) = -\frac{2}{1-e^{-\beta\Omega}}  \im \chi(\Omega).
\end{align}
% Since we are  interested in the Raman scattering a
At finite temperatures, we will work in
 the Matsubara formalism, in which the  Matsubara correlation function of Raman operators is given by $
\mI(\tau) =  -\left\langle T_\tau \mR(\tau) \mR(0)  \right\rangle $ and the Fourier transform
 can be written as
 \begin{align}
 \mI(i\Omega_n) =  -\int \ud \tau e^{i\Omega_n \tau}\left\langle T_\tau \mR(\tau) \mR(0)  \right\rangle.
 \end{align}
 After analytical continuation $i\Omega_n \to \Omega + i\delta $, we directly obtain the retarded correlation function, i.e. the Raman susceptibility, 
\begin{align}
    \chi(\Omega)=\mI({i\Omega_n})|_{i\Omega_n\rightarrow\Omega+i\delta }.
\end{align}
 So in the following derivation, we only need to focus on evaluating the Matsubara correlation function of the Raman operators $\mI(i\Omega_n)$.

Applying this mechanism, we first compute the Raman response of the decoupled phonon and spin subsystems, in which case $\mI_0 =  \mI_\tx{em-ph} + \mI_\tx{em-s}$. The first term describes the pure phonon Raman scattering:
\begin{align}
 \mI_\tx{em-ph}^{\mu\mu'}(i\Omega_n) = \sum_{\Gamma m} \mu^2_\Gamma  \left(R_{\Gamma m}^{\mu\mu'}\right)^2    \mD^{(0)}_{\Gamma\Gamma, m m}(i\Omega_n)
\end{align}
where the scattering geometry $\mu\mu'$ has been explicitly specified, $\mu_\Gamma$  is the photon-phonon coupling constant in the $\Gamma$ irrep and $R_{\Gamma m}^{\mu\mu'}$ is  the Raman polarization tensor defined by \refeq{Ramantensor}
and 
$
    \mD^{(0)}_{\Gamma m, \Gamma'm'}(i\omega_n)=
    \frac{2\omega_{\Gamma} }{\left(i \omega_{n}\right)^{2}-\omega_{\Gamma}^{2}  }   \delta_{\Gamma\Gamma'} \delta_{mm'}
$
is the bare phonon propagator. 
The corresponding Raman response  is simply given by a set of delta functions at the bare phonon frequencies $\omega_{\Gamma}$. 

The second term comes from the  magnetic Raman scattering:
\begin{align}
 \mI_\tx{em-s}^{\mu\mu'}(i\Omega_n) =  -\int_0^\beta \ud \tau e^{i\Omega_n \tau}\langle T_\tau \mR_\tx{em-s}^{\mu\mu'}(\tau)\mR_\tx{em-s}^{\mu\mu'}(0)\rangle.
 \end{align}
%  where $ {\mathcal R}^{\mu\mu'}_{\tx{em-s}} =\nu \sum_{m}  \tilde{ R}^{\mu\mu'}_{E_g,m} \Sigma_{E_g,m} $ is given in the main text and $\tilde{ R}^{\mu\mu'}_{E_g,m}$ defined in Eq.(B5).
 The spin bilinear operator can be rewritten using the Majorana fermion representation of the spin: $\sigma_j^\alpha = ib_j^\alpha c_j$, and then transformed into the basis of the fermionic eigenmodes  \cite{Feng2021}.
  Explicitly, the  correlation function of the spin Raman operators is written as a general form of $
 - \left\langle T_{\tau}\left(\mathbf{B}^{\dagger} \tilde{\Lambda} \mathbf{B}\right)(\tau)\left(\mathbf{B}^{\dagger} \tilde{\Lambda} \mathbf{B}\right)(0)\right\rangle 
$, where  $\mathbf{B}^{\dagger}=\left[\beta_{1}^{\dagger},\cdots \beta_{N}^{\dagger}, \beta_{1},\cdots \beta_{N}\right]$  is the vector of the Bogoliubov quasiparticles,  and 
$\tilde{\Lambda}$ is a symmetrized coupling matrix, whose entries are the coupling vertices between two fermion eigenmodes and the photons.  Since the fermionic eigenmodes are different for different flux configurations,
coupling matrix $\tilde{\Lambda}$ is a function of $Z_2$ gauge fluxes. For a given temperature, the thermodynamic average over different flux configurations is evaluated by {\em stratified Monte Carlo (strMC)} method, which was  developed and applied on acoustic phonon dynamics simulations for the Kitaev QSL system in Ref.~\cite{Feng2021,feng2022phonon}.

Then,  $\mI_\tx{em-s}^{\mu\mu'}(i\Omega_n) $ is evaluated as
\begin{align}
\mI_\tx{em-s}^{\mu\mu'}(i\Omega_n) \sim \Tr \left[ {\mathcal G}_{1}(i\omega_{n}) \tilde{\Lambda} {\mathcal G}_{1}^{*}(i\Omega_n - i\omega_{n_1}) \tilde{\Lambda} +{\mathcal G}_{2} (i\omega_{n})\tilde{\Lambda} {\mathcal G}_{2}(i\Omega_n - i\omega_{n_1})\tilde{\Lambda}^\T \right],
\end{align}
which appears as a fermionic loop diagram shown in \reffg{Fig_Feynman}(b). Here,
the indices $\mu\mu'$ are contained inside $\tilde{\Lambda}$, ${\Tr} [...]$ sums over the Matsubara frequencies $i\omega_{n_1}$ as $T\sum_{n_1}$, and
 the matrix form of the 
 Matsubara Green's functions is given by:
\begin{align}
\mathcal{G}_{1}\left(i \omega_{n}\right) & \equiv\left[\begin{array}{ll}
\bar{g}\left(i \omega_{n}\right) & O \\
O & g\left(i \omega_{n}\right)
\end{array}\right],
\\
\mathcal{G}_{2}\left(i \omega_{n}\right) & \equiv\left[\begin{array}{ll}
O & g\left(i \omega_{n}\right) \\
\bar{g}\left(i \omega_{n}\right) & O
\end{array}\right],
\end{align}
where, 
$g_{i}\left(i \omega_{n}\right)=\frac{1}{i \omega_{n}-\epsilon_{i}}$ and 
$\bar{g}_{i}\left(i \omega_{n}\right)=\frac{1}{i \omega_{n}+\epsilon_{i}}$.
%denote the Matsubara Green's function for each fermionic mode.
%In the following, all fermionic loops
%, eg. the polarization bubble $\Pi$ and the spin-dependent phonon Raman vertex $\mP_L, \mP_R$, 
%will be evaluated in the same way. 
The spectrum of $\mI_\tx{em-s}^{\mu\mu'}$ appears as a magnetic continuum, which has been studied at length in the literature \cite{Knolle2014, Nasu2016,Rousochatzakis2019} and will not be repeated here.
 
Now, we are ready to turn on the spin-phonon coupling, which is the main goal of this work. We will derive the formula for the Raman response for 
the spin-phonon coupled Kitaev system described by the Hamiltonian \refeq{eq:model}.
The presence of the spin-phonon interaction (\refeq{eq: Hsp} in the main text)  leads to the  Raman vertex renormalization due to the final-state interactions.
% The origin of the magnetoelastic coupling leads to creation either a phonon  in the intermediate state of the magnetic Raman scattering (c) or, equivalently, excitation of a pair Majorana fermions in the intermediate state of the phonon Raman scattering.
In the interaction picture, the general expression of the Raman correlation function in the presence of the spin-phonon coupling is given  by
\begin{align}
I(t) =  \langle T_t \mathcal{R}(t)\mathcal{R}(0)  e^{-i \int \ud t' H_\tx{s-ph}(t')} \rangle,
\end{align}
where  $S= e^{-i \int \ud t' H_\tx{s-ph}(t')}$ is the dubbed  $S$-matrix. Correspondingly, at finite temperature $ \mI(\tau) =  -\left\langle T_\tau \mR(\tau) \mR(0) e^{-\int_0^\beta \ud \tau' H_\tx{s-ph}(\tau')}  \right\rangle$ gives the Matsubara correlation function of the Raman operator in the spin-phonon coupled Kitaev model. Treating the coupling $H_\tx{s-ph}$  perturbatively  and using 
the $S$-matrix expansion  \cite{Mahan}, we obtain:
\begin{align}
 \mI(\tau) = -\sum_{k=0}^\infty (-1)^k  \prod_{i}^k \int_0^\beta \ud\tau_i \lefta T_\tau 
     \mathcal{R}(\tau)\mathcal{R}(0)
     \prod_{i}^k H_\tx{s-ph}(\tau_i) \righta \label{eq: Sexpansion}
\end{align}
%{\it [There are indeed two $\prod$ here]}
where only connected different graphs are summed.  
 At the order of $k=0$, this expression corresponds to the simple spin-phonon decoupled case, as described above. 
\begin{figure}
 \includegraphics[width=0.8\columnwidth]{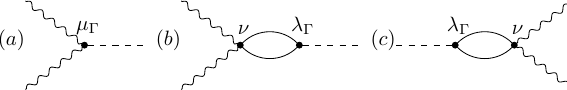}
 \caption{
 The Feynman diagrams of the phonon Raman vertices: (a) $\mu_\Gamma R_{\Gamma m}$ (b) $\mP_{\Gamma m, L}$ (c) $\mP_{\Gamma m, R}$.
}
 \label{Fig_vertices}
 \end{figure}

\begin{figure}
 \includegraphics[width=0.75\columnwidth]{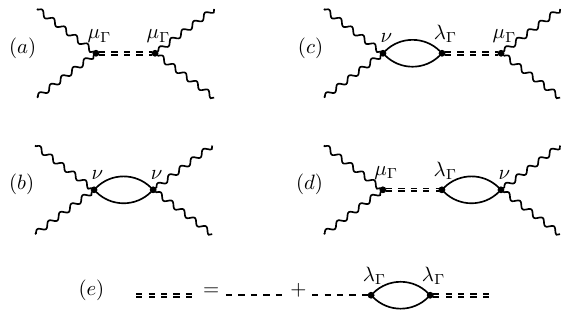}%
 \caption{
 The Feynman diagrams of the Raman intensity shown in \refeq{eq: Raman_inten_app}. 
 (a) the phonon channel with a propagator renormalized by the  spin-phonon interaction (\refeq{eq: Hsp} in the main text), (b) the spin channel,
  (c)-(d) the phonon-spin mixed channel with the spin-dependent phonon Raman vertices $\mP_{\Gamma m, L}$, $\mP_{\Gamma m, R}$.
  Panel (e) shows the Dyson's equation   for the phonon propagator.
}
 \label{Fig_Feynman}
 \end{figure}

At the order of $k=1$, contribution can be explicitly written as
\begin{align}
 \mI_1^{\mu\mu'}(\tau)&= \int_0^\beta \ud\tau_1 \lefta T_\tau 
     \mathcal{R}^{\mu\mu'}(\tau)\mathcal{R}^{\mu\mu'}(0)
     H_\tx{s-ph}(\tau_1) \righta =\left\{
     \begin{array}{l}
     \int_0^\beta \ud\tau_1 \lefta T_\tau 
     \mathcal{R}^{\mu\mu'}_\tx{em-s}(\tau)\mathcal{R}^{\mu\mu'}_\tx{em-ph}(0)
     H_\tx{s-ph}(\tau_1) \righta\\
     \int_0^\beta \ud\tau_1 \lefta T_\tau 
     \mathcal{R}_\tx{em-ph}^{\mu\mu'}(\tau)\mathcal{R}^{\mu\mu'}_\tx{em-s}(0)
     H_\tx{s-ph}(\tau_1) \righta
     \end{array}
     \right.\nonumber\\
     &=
     \left\{
     \begin{array}{l}
    \sum_{\Gamma,m,m'} \mu_\Gamma \lambda_\Gamma R_{\Gamma m}^{\mu\mu'}  \int_0^\beta \ud\tau_1 \lefta T_\tau 
       \mathcal{R}^{\mu\mu'}_\tx{em-s}(\tau)
     \Sigma_{\Gamma m'} (\tau_1)u_{\Gamma m'} (\tau_1) u_{\Gamma m} (0)
     \righta\\
    \sum_{\Gamma,m,m'} \mu_\Gamma \lambda_\Gamma R_{\Gamma m}^{\mu\mu'} \int_0^\beta \ud\tau_1 \lefta T_\tau 
      u_{\Gamma m}
     (\tau)
     \Sigma_{\Gamma m'}(\tau_1) u_{\Gamma m'} (\tau_1) \mathcal{R}_\tx{em-s}^{\mu\mu'}(0)\righta
     \end{array}
     \right. \label{I1}
\end{align}
which contributes into the lowest order of the diagrams shown in (c)(d) respectively  \reffg{Fig_Feynman} (c-d).
At this order, it gives to the {\it spin-dependent} phonon Raman vertices (first introduced in \cite{Moriya1967, Suzuki1973}), which describe the mixing term between the two channels and  play the central role in generating the Fano lineshape. These two spin-dependent phonon Raman vertices, which are distinguished with notation left (L) and right (R), are given by
\begin{align}
    \mP^{\mu\mu'}_{\Gamma m, L} (\tau) = - \lambda_\Gamma \lefta T_\tau   \Sigma_{\Gamma m} (\tau)  \mR^{\mu\mu'}_\tx{em-s} (0) \righta,  \nn
\\
    \mP^{\mu\mu'}_{\Gamma m, R} (\tau) = - \lambda_\Gamma \lefta T_\tau  \mR^{\mu\mu'}_\tx{em-s} (\tau)  \Sigma_{\Gamma m} (0) \righta.
\end{align}
The corresponding diagrams are shown in \reffg{Fig_vertices} (b-c).
Combining $\mP^{\mu\mu'}_{\Gamma m, L} (\tau)$  and $\mP^{\mu\mu'}_{\Gamma m, R} (\tau)$  with the bare Raman vertex $ R_{\Gamma m}^{\mu\mu'}$, 
we define the renormalized left and right phonon Raman vertices as:
\begin{align}
    {R'}_{\Gamma m, L}^{\mu\mu'}(\tau) = \mu_\Gamma  R_{\Gamma m}^{\mu\mu'} +  \mP^{\mu\mu'}_{\Gamma m, L}(\tau), \nn
    \\
    {R'}_{\Gamma m, R}^{\mu\mu'}(\tau) = \mu_\Gamma R_{\Gamma m}^{\mu\mu'} +  \mP^{\mu\mu'}_{\Gamma m, R}(\tau). \label{eq: total_Raman}
\end{align}
 With these renormalized phonon Raman vertices, the odd-$k$ terms and even-$k$ terms in the expansion \refeq{eq: Sexpansion} are grouped together, and the summation naturally forms the series that is consistent with Dyson's equation, which describes the renormalization of the phonon propagator:
\begin{align}
\mD_{\Gamma m, \Gamma' m'} = \leftr [\mD^{(0)}_{\Gamma m, \Gamma' m'}]^{-1} - \Pi_{\Gamma m, \Gamma' m'} \rightr^{-1} \label{eq: Dyson}
\end{align}
described by \reffg{Fig_Feynman} (e).
Here, $
    \Pi_{\Gamma m, \Gamma' m'}(\tau) = -\lambda_\Gamma\lambda_{\Gamma'} \lefta T_\tau \Sigma_{\Gamma m} (\tau) \Sigma_{\Gamma'm'}(0) \righta   \label{eq: Pi}
$
is the polarization bubble given by \refeq{eq: pol_bubble} of the main text. Its temperature and field dependence are discussed in the section.
Then the final expression of the Raman correlation function can be obtained: 
\begin{align}
    \mI (\tau)= \mI_\tx{em-s} (\tau) + R'_{L} (\tau) \cdot \hat{\mD}  (\tau)\cdot R'_{ R} (\tau), \label{eq: Raman_inten_app}
\end{align}
where the dot product is on the contraction of ($\Gamma, m$) indices. This result is summarized in Fig.~\ref{Fig_Feynman} (a)-(d), where the diagrams (a),(c) and (d) are contained in the second term of the above expression.
The renormalized phonon Raman vertices \refeq{eq: total_Raman} can be  also written as
$R'_{\Gamma m}(\tau) = R_{\Gamma m} \left[\mu_\Gamma + \frac{\nu}{\lambda_\Gamma} \Pi_{\Gamma m, \Gamma m}(\tau)\right]$, when $\kappa=0$. 
%where $\mathcal{C}_\Gamma(\tau)\equiv\mathcal{C}_{E_g}(\tau) = -\lefta T_\tau \Sigma_{{E_g} m}(\tau) \Sigma_{{E_g} m}(0) \righta $.
%is the spin bilinear operator correlation function, and also corresponds to a fermionic loop diagram.
In the frequency domain, the fermionic bubble gives the {\it frequency-dependent} renormalization of the phonon Raman coupling $\mu_\Gamma$, which eventually leads to the asymmetry of  the phonon Raman peak.

\setcounter{equation}{0}
\renewcommand{\thesection}{D}
 \section{D. The polarization bubble and its influence on the phonon peaks}
 \label{sec:D}
 In this section, we will analyze the temperature and field dependence of the polarization bubble $
    \Pi_{\Gamma m, \Gamma' m'} = -\lambda_\Gamma\lambda_{\Gamma' }\lefta T_\tau \Sigma_{\Gamma m} (\tau) \Sigma_{\Gamma' m'}(0) \righta,
$ and  discuss its effects on the shape of the phonon Raman peaks. 
%The polarization bubble
%$\Pi_{\Gamma m, \Gamma' m'} &= -\lambda_\Gamma\lambda_{\Gamma' }\lefta T_\tau \Sigma_{\Gamma m} (\tau) \Sigma_{\Gamma' m'}(0) \righta,
%$
%renormalizes the free phonon propagator according to the Dyson's equation (\ref{eq: Dyson}). 
As the $E_g^1$ and $E_g^2$ phonon peaks are energetically well separated \cite{Li2019},
the off-diagonal components of  $\Pi_{\Gamma m, \Gamma' m'}$ are negligible. Thus, we will focus only on the diagonal blocks of $\hat{\Pi}$, which are denoted as $\Pi_{m m '} \equiv \Pi_{\Gamma m, \Gamma m'}$.

\begin{figure}
 \includegraphics[width=0.8\columnwidth]{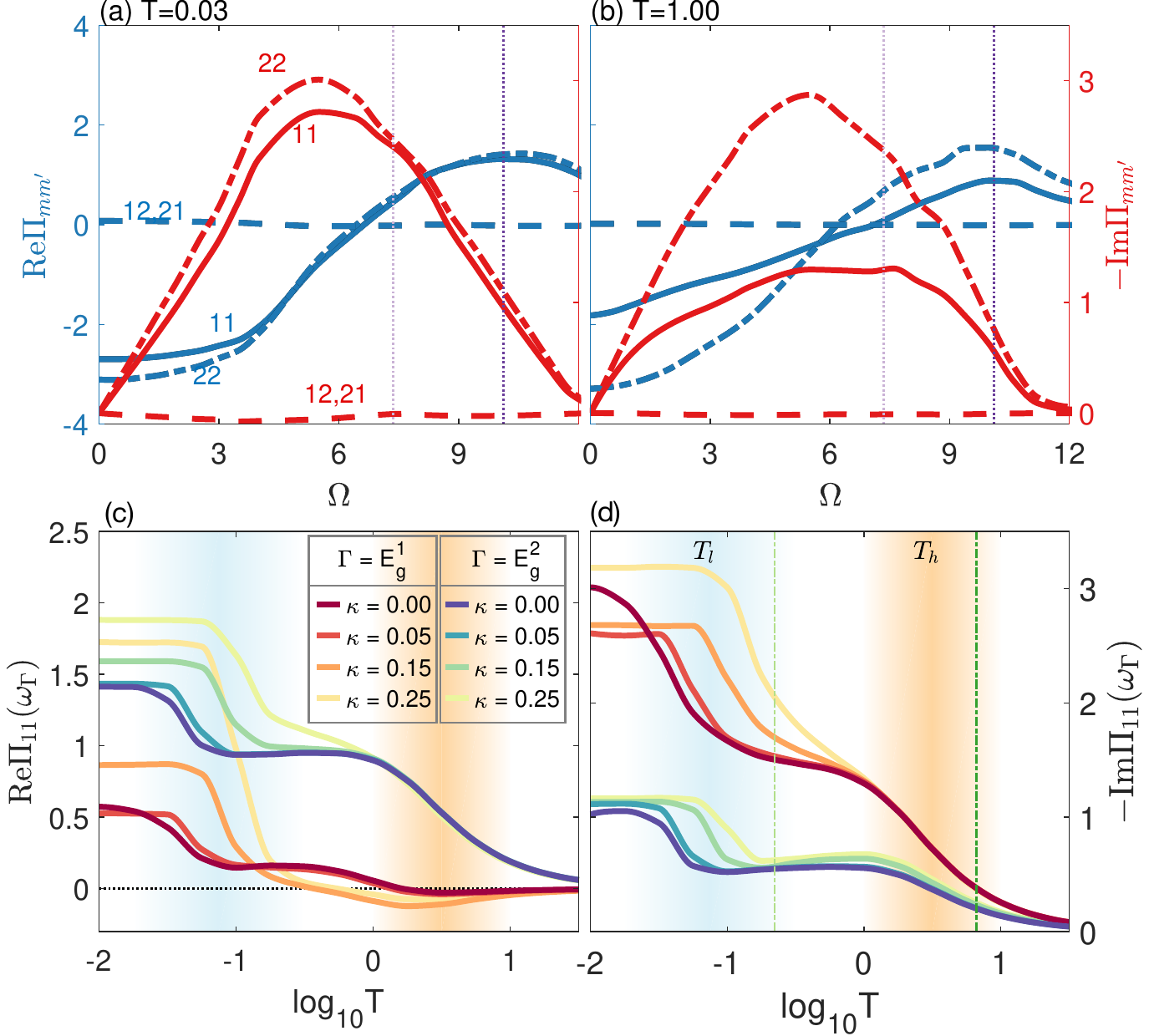}%
 \caption{The real and imaginary part of the polarization bubble $\Pi_{mm'}$  within a $E_g$ channel measured in the units of $\lambda^2_\Gamma$ ($mm'$ components have been indicated in (a)).
Panels (a-b): the frequency dependence of $\re\Pi_{mm'}$ and  $-\im\Pi_{mm'}$ 
at different temperatures and $\kappa=0$, corresponding to average flux densities of (a) $n_\tx{av}=0.01$, (b) $n_\tx{av}=0.48$.
 The two vertical purple lines denote the bare phonon energies $ \omega_{E^1_{g}} = 7.32$   and  $ \omega_{E^2_{g}} =10.1$. 
 Panels (c-d): the temperature dependence of $\re\Pi_{11}$ and $-\im\Pi_{11}$ for various $\kappa$ evaluated at $\omega_{E^1_{g}}$ and $\omega_{E^2_{g}}$ respectively.
$T_l$ and $T_h$ are the two crossover temperatures. The two green vertical dashed lines in (d) indicate $T=5$ K  and 150 K. All results are obtained by the strMC method.}
 \label{Fig2}
 \end{figure}

In \reffg{Fig2}, we present the  real  and imaginary parts of  $\Pi_{m m'}$ 
(blue and red curves, respectively) as functions of frequency and temperature, computed by strMC method \cite{Feng2021, feng2022phonon}. 
\reffg{Fig2}(a-b) show the components of  $\Pi_{m m'}$  computed  for $\kappa=0$ at temperatures $T=0.03$ and $T=1$, respectively.
We can see that both at   low ($T=0.03$) and high ($T=1$) temperatures
$\Pi_{12}$ and $\Pi_{21}$ are negligibly small, which shows that the two degenerate phonon modes are indeed orthogonal.
Moreover, 
both $\re \Pi_{11}$ (blue  solid curve) and $\re \Pi_{22}$ (blue dot-dash curve)  are positive when evaluated at both $\omega_{E_g^1}$ and $\omega_{E_g^2}$ phonon energies, which indicates that the renormalized phonon energies are larger than bare phonon energies. This remains qualitatively unchanged even at high temperature when $Z_2$ fluxes proliferate.
However, the temperature evolution of $\Pi_{mm'}$  shows the quantitative difference between 11 and 22 components: the 11 component is more sensitive to the thermal flux disorder. This difference is solely determined by the specific form of the spin irreducible representations $\Sigma_{\Gamma m}$ as introduced in the main text \refeq{eq: Hsp}. 
On the other hand,
the imaginary part $-\im \Pi_{mm'}$ engenders the finite phonon life-time, which gives rise to an increase of the phonon's peak width.

 We focus on the $xx$-scattering geometry, which is the mostly used in the experiment.
  As  shown in   \refeq{Ramantensor}, the Raman tensor $ R_{E_g,1}^{\mu\mu'}$ ($m=1$) has nonzero $xx$  and  $yy$ diagonal components (corresponding to the parallel polarization),  while $ R_{E_g,2}^{\mu\mu'}$ ($m=2$)  has only  off-diagonal components (corresponding to the cross polarization). Therefore, the renormalization of the position and peak's width in the   $xx$-scattering geometry are mainly controlled by $\Pi_{11}$.
 In \reffg{Fig2}(c-d), we show the temperature dependence of  $\Pi_{11}(\omega_{\Gamma})$  computed at bare phonon frequencies 
$\omega_{\Gamma}=\omega_{E_g^1}$ and  $\omega_{\Gamma}=\omega_{E_g^2}$
for various values of $\kappa$ (recall that $\kappa$  mimics the effect of an external magnetic field).
 We can see that both $\re \Pi_{11}(\omega_{\Gamma})$ and $\im \Pi_{11}(\omega_{\Gamma})$ display  a two-stage decrease with increasing temperature, which is shared by other thermodynamics quantities in the Kitaev spin liquid \cite{Nasu2015,Feng2020,Feng2021}. The two crossover temperatures, namely $T_l$ (in blue shaded area) and $T_h$ (in orange shaded area) correspond, respectively, to the flux proliferation temperature and the major fermionic excitation temperature. 
 While $T_h$ is almost insensitive to
 $\kappa$, $T_l$ increases with $\kappa$, which results from the increase of the $Z_2$ flux gap energy \cite{Feng2021, Feng2020}.
 
 We now can perform an explicit calculation of the  Raman phonon lineshape.
Based on the Dyson equation (\ref{eq: Dyson}), the renormalization of the phonon energy and broadening  of the peak's width can be estimated from the polarization bubble $\hat{\Pi}$. 
When $\kappa= 0$, the off-diagonal components of the polarization bubble $\Pi_{mm'}$ is negligible as shown in \reffg{Fig2}.
So the imaginary part of a diagonal entry of the renormalized phonon propagator is given by (we have explicitly 
moved $\lambda_\Gamma$ out of $\Pi$):
\begin{align}
    -\im D_{mm}(\Omega) = 
    \frac{4\omega^2_\Gamma 
         \left(
            \frac{ \Omega\delta_{\tx{ph}}}{\omega_\Gamma} - \lambda_\Gamma^2 \im\Pi_{mm} 
        \right)}
    { \left(
        \Omega^2 - \omega_\Gamma^2 - 2\omega_\Gamma\lambda_\Gamma^2 \re\Pi_{mm}
    \right)^2 +
    4\omega_\Gamma^2\left(
       \frac{ \Omega\delta_{\tx{ph}}}{\omega_\Gamma} - \lambda_\Gamma^2 \im\Pi_{mm} 
    \right)^2 }, \label{eq: D1}
\end{align}
where the analytical continuation $i\Omega_n \to \Omega + i\delta_{\tx{ph}}$ has been taken to obtain retarded correlation function $D_{mm} (\Omega)$, and $\delta_{\tx{ph}}$ is an artificial broadening of the bare phonon peak.
As mentioned above, the xx-geometry scattering is controlled by the $mm=11$ component. Then, the half width at half maxima (HWHM) of the phonon peak can be estimated as
\begin{align}
    \gamma_\tx{est} &=  - \lambda_{\Gamma}^2 \im\Pi_{11} + \delta_{\tx{ph}}. \label{eq: estim0}
\end{align} 
 Here, note that there is an artificial background contribution to the line width, $\delta_\tx{ph}$, which causes the nonzero line width at $T=\infty$ (numerically at $T=10^{1.5}$). Therefore, both in the main text and here, we offset the computed line width by a background value obtained at the infinite temperature.
% {\cred Here, we note that the artificial background contribution to the line width $\delta_\tx{ph}$, or  an offset  for the line width,   is taken to be the  same at all temperatures, including   $T=\infty$ (in practice corresponding  to  $T=10^{1.5}$).
As shown in Tab.~\ref{tab: Gamma}, the decrease of peak width $\gamma_\tx{MC}$ (evaluated by the strMC simulations)
 between 5 K and 150 K is estimated to be $0.055J$. This is comparable to the experimental findings in Ref.~\cite{Sandilands2015}, that the anomalous peak width $\gamma_\tx{exp}$ displays a decrease of $0.085 J$  between 5 K and 150 K .
This result indicates that the source of the anomaly comes from the spin-phonon coupling in the vicinity of the Kitaev spin liquid. 
In the low-temperature region between 0.2 K to 5 K, the estimation  of the peak width differs significantly from the strMC result. This is due to the effect of the spin-dependent phonon Raman coupling at low temperatures also causes smaller peak width.

The renormalized peak's position, $\omega^{\tx{ren}}_\Gamma$, is given by
\begin{align}
    \omega^{\tx{ren}}_\Gamma &= \sqrt{\omega_\Gamma^2 + 2\omega_\Gamma\lambda_\Gamma^2\re\Pi_{mm}} \approx \omega_\Gamma + \lambda_\Gamma^2  \re\Pi_{mm}. \label{eq: estim_omg}
\end{align} 
 Since   $\re\Pi_{11} >0$,  the phonon peak moves towards right.  This energy shift decreases with increasing temperature.

 \begin{table}
    \begin{center}
    \begin{tabular}{l|cccccccc}
    \hline
    $T/K$ & \fsz 0.2 & \fsz 5  & \fsz 150 & \fsz $\infty$  \\ 
    \hline
    $\gamma_{\tx{est}}/J$ & \fsz 0.280 & \fsz 0.150  & \fsz 0.055  & \fsz 0  \\
    \hline
    $\gamma_{\tx{MC}}/J$ & \fsz 0.225 & \fsz 0.090 & \fsz 0.035 & \fsz 0  \\
    \hline
    $\gamma_{\tx{exp}}/J$ &  & \fsz 0.090 & \fsz 0.005 &   \\
    \hline
    \end{tabular}
     \caption{The half width at half maxima (HWHM) of the $E_g^2$ phonon peak at several characteristic temperatures.  $\gamma_{\tx{est}}$ is the HWHM estimated as $\gamma_{\tx{est}}=-\lambda_{E_g^2}^2\im\Pi_{11}$, where $\lambda_{E_g^2} = 0.52$ (in units of $J$).
    $\gamma_{\tx{MC}}$ is HWHM of $E_g^2$ phonon obtained from \refeq{eq: Raman_inten_app}, evaluated with the strMC simulations on a lattice of $N_1=N_2 = 25$ as shown in \reffg{fig: FullModelFano}(b) in the main text. Note that $\gamma_\tx{MC}$ here has been offset by a background line width obtained at $T=\infty$, which mainly comes from the artificial broadening $\delta_\tx{ph}$.
   $\gamma_{\tx{exp}}$ is the experimentally measured HWHM, which is the anomalous deviation from the anharmonic behaviour. It is obtained from Ref.~\cite{Sandilands2015}, and $\gamma_{\tx{exp}} = \Delta \Gamma/2$ in the notations therein. The two temperatures $5K$ and $150K$ are marked as green dot-dashed lines in \reffg{fig: FullModelFano}(b) in the main text.
    }
  \label{tab: Gamma}
  \end{center}
\end{table}

\setcounter{equation}{0}
\renewcommand{\thesection}{E}
\section{E. The details of the model fitting}

In this section, we describe the details of fitting  the experimental Raman spectrum and clarify the uniqueness of the best-fit model parameters $\{\omega_\Gamma,\lambda_\Gamma, \nu, \mu_\Gamma\}$ up to an overall scaling.
First we note that the overall magnitude of Raman spectrum is free to rescale. This degree of freedom is reflected in a simultaneous scaling of the couplings $\{\nu, \mu_\Gamma\}$. If they are magnified or reduced uniformly by the same factor, then the resultant calculated spectrum will retain its shape with only an overall scale difference. This can be clearly seen from \refeq{eq: Raman_inten_app} and Feynman diagrams presented in \reffg{Fig_Feynman}. Thus, we can set $\mu_{E_g^2} = 1$ to fix the overall scale, and  get $\mu_{E_g^1} = 0.36 \mu_{E_g^2}$ and $\nu = -0.63 \mu_{E_g^2}$. After getting rid of the overall scaling factor, the model parameters are uniquely decided by the process of fitting the computed Raman intensity to the experimental Raman curve.
In this process, $\nu$ controls the overall intensity of the magnetic continuum which is contributed from both $I_\tx{em-s}$ and the spin-dependent phonons Raman couplings, and it  also affects the Fano asymmetry of  both phonon peaks.
$\lambda_\Gamma$ controls the phonon peak widths as well as the Fano asymmetry of respective peaks.
$\mu_\Gamma$ controls the phonon peak heights. $\omega_\Gamma$ controls the peak positions. Therefore, each parameter has its unique effect, and changing one of these parameters will not be completely compensated by tuning the others. This guarantees that the optimal set of the model parameters is unique.

% \setcounter{equation}{0}
% \renewcommand{\thesection}{F}
% \section{F. Validation of the spin-phonon coupling constant.} 

% In the main text, an estimation of spin-phonon coupling constants $\lambda_\Gamma$ are obtained by fitting the experimental Raman spectrum. But the estimation $\lambda_{E_g^2} = 1.3 J$ is at the same order of $J$, so we need to check the validity of the perturbation calculation. 

% First of all, the spin-phonon coupling $\lambda_{E_g^2} = 1.3 J$ is still small compared to the two bare phonon energies considered in our paper. So its renormalization to the phonon dynamics is still consistent with the perturbative calculation. This renormalization is explicitly shown in \ref{eq: D1}. 

\setcounter{equation}{0}
\renewcommand{\thesection}{F}
\section{F. Absence of the Fano lineshape in  the phonon Raman response with perpendicular polarization.} 

In this section, we apply our theory to analyze the polarization-resolved Raman experiment in $\alpha$-RuCl$_3$ reported in Ref.~\cite{Mai2019}. This work explores the Raman spectroscopy of the out-of-plane polarizations, and concludes that the spin-related effects, namely the magnetic continuum and Fano lineshape asymmetry, disappear when the photon polarization is perpendicular to the honeycomb plane of  Ru$^{3+}$ ions, suggesting  that these effects are  both of the same two-dimensional origin. 
% {\cbl Here, we will show that the nonzero phonon Raman scattering for perpendicular polarization is a result of symmetry breaking from $D_{3d}$ to $C_{2h}$; also, this experiment provides extra constraint on the form of the spin-photon coupling.}

To explore the polarization dependence of the Raman spectroscopy, \refeq{eq: Raman_inten_app} needs to be explicitly evaluated. Following the same set up in Ref.~\cite{Li2019}, for polarization within the $a$-$c$ plane, we denote the angle between $E$ and $a$ axis as $\phi$. Consider parallel scattering geometry, $E^\mu_{in} = E^\mu_{out} = [\cos \phi, 0, \sin\phi]$. Then the Raman intensity is proportional to 
\begin{align}
    \cos^4\phi\, \mI^{xx} + \sin^4\phi\, \mI^{zz} + \cos^2\phi\sin^2\phi\, \mI^{xz} + \sin^2\phi\cos^2\phi\, \mI^{zx},
\end{align}
where $\mI^{\mu\mu'}$ is defined in \refeq{eq: Raman_inten_app}. We focus on the perpendicular polarization $\phi=\pi/2$ so only $\mI^{zz}$ is considered. First we analyze the phonon peak in the Raman spectrum, which is contributed from the phonon Raman tensor $R_{\Gamma m}^{\mu\mu'}$. As shown in \refeq{Ramantensor}, $R_{\Gamma m}^{zz} = 0$ for $D_{3d}$ group, which indicates that the response  in the $zz$ polarization would be identically zero.
But if the symmetry is broken to $C_{2h}$ point group due to the monoclinic distortion in real $\alpha$-RuCl$_3$ crystals \cite{Li2019, johnson2015monoclinic, cao2016low}, the non-zero $R_{\Gamma m}^{zz}$ is allowed so the phonon peak persists at $\phi=0$ there. Therefore, the nonzero peak in the Raman spectrum at perpendicular polarization observed in Ref.~\cite{Mai2019} must result from the  symmetry breaking from $D_{3d}$ to $C_{2h}$. This symmetry breaking is introduced from the distortion of the honeycomb lattice due to the weak interlayer interaction \cite{Mai2019}.
% Second, the spin-related effects, namely the magnetic continuum and Fano effect, are respectively from $\mI_{sR}^{\mu\nu}$ (\ref{eq: sR}) and $\rho_{rm,L(R)}^{\mu\nu}$ (\ref{eq: rho}), both of which are proportional to $\Sigma_{LF}^{\mu\nu}$ (\ref{eq: Sig_LF}). Since $\Sigma_{LF}^{zz} = 0$, the spin-related effects disappear for perpendicular polarization, which explains the 2D effect observed in Ref.~\cite{li2019raman}. But note, again, that in a more comprehensive study \cite{Yang}, non-LF terms also contribute to spin-Raman coupling. Here we have assumed that $\Delta H_{sR} = \nu\Sigma_{LF}^{\mu\nu}E^{\mu}E^{\nu}$ is a valid approximation for Raman spectroscopy at the energy scale above $1J$, which is validated by the results in Ref.~\cite{knolle2014raman}. Conversely, the analysis here also shows that, to produce the 2D effects as in Ref.~\cite{li2019raman}, the spin Raman coupling Hamiltonian $\Delta H_{sR}$ should have negligible zz component, which offers a constraint on the theoretical modelling of this coupling Hamiltonian.

%   Therefore, the nonzero peak in the Raman spectrum at perpendicular polarization  observed in Ref. \cite{Mai2019} must result from the  symmetry breaking from $D_{3d}$ to $C_{2h}$,  and deviation of the perpendicular direction $c$ from the cubic $z$-axis.  However, since the angle between  $c$  and  $z$ is very small, the coupling to the magnetic continuum is weak and thus almost no asymetry characteristic to the Fano  lineshape was observed in \cite{Mai2019}. 

\end{document}